\documentclass[11pt,a4paper]{article}
\usepackage[colorlinks=true, linkcolor=black!50!blue, urlcolor=blue, citecolor=blue, anchorcolor=blue]{hyperref}
\usepackage[font=small,labelfont=bf,margin=0mm,labelsep=period,tableposition=top]{caption}
\usepackage[a4paper,top=2cm,bottom=2.5cm,left=2cm,right=2cm,bindingoffset=0mm]{geometry}

\usepackage{graphicx}
\usepackage{float}
\usepackage{afterpage}
\usepackage{epsfig,cite}
\usepackage{amssymb}
\usepackage{amsmath}
\usepackage{dsfont}
\usepackage{pdflscape}
\usepackage{multirow}
\usepackage{url,hyperref}
\usepackage{booktabs}
\usepackage{tabularx}
\usepackage{xcolor}
\usepackage{bm}
\usepackage{textcomp}
\usepackage{caption}
\begin{document}

\vspace{-2.0cm}
\begin{flushright}
\end{flushright}
\vspace{.1cm}

\begin{center} 

 {\Large \bf Helicity-dependent parton distribution functions\\
    \vspace{0.2cm}at next-to-next-to-leading order accuracy\\
  \vspace{0.2cm}from inclusive and semi-inclusive deep-inelastic scattering data\\
  \vspace{.5cm}
}

Valerio~Bertone$^a$, Amedeo Chiefa$^{b,c}$, and Emanuele R. Nocera$^c$\\
\vspace{.3cm}
(MAP Collaboration)\footnote{The MAP acronym stands for ``Multi-dimensional
  Analyses of Partonic distributions''. It refers to a collaboration aimed at
  studying the three-dimensional structure of hadrons.}\\
\vspace{.3cm}
{\it ~$^a$ IRFU, CEA, Universit\'e Paris-Saclay, F-91191
  Gif-sur-Yvette, France}\\
{\it ~$^b$ The Higgs Centre for Theoretical Physics, University of Edinburgh,\\
  JCMB, KB, Mayfield Rd, Edinburgh EH9 3JZ, Scotland}\\
{\it ~$^c$ Universit\`a degli Studi di Torino and INFN, Torino,\\
  Via P.~Giuria 1 I-10125 Torino, Italy}\\

\end{center}   

\vspace{0.1cm}

\begin{center}
  {\bf \large Abstract}\\
\end{center}
  
We present {\sc MAPPDFpol1.0}, a new determination of the
helicity-dependent parton distribution functions (PDFs) of the proton
from a set of longitudinally polarised inclusive and semi-inclusive
deep-inelastic scattering data. The determination includes
next-to-next-to-leading order QCD corrections to both
processes, and is carried out in a framework that combines a
neural-network parametrisation of PDFs with a Monte Carlo
representation of their uncertainties. We discuss the quality of the
determination, in particular its dependence on higher-order
corrections, on the choice of data set, and on theoretical
constraints.

\vspace{1cm}

\section{Introduction}
\label{sec:intro}

The accurate and precise determination of helicity-dependent (polarised,
henceforth) proton parton distribution functions (PDFs)~\cite{Ethier:2020way}
is key to understand the quark and gluon spin structure of the
nucleon~\cite{Ji:2020ena}. This endeavour, which started more than thirty years
ago with the first measurements of the spin-dependent structure function $g_1$
in polarised inclusive deep-inelastic scattering (DIS) by the European Muon
Collaboration~\cite{EuropeanMuon:1987isl,EuropeanMuon:1989yki}, is
the focus of the forthcoming Electron-Ion Collider
(EIC) experimental program~\cite{AbdulKhalek:2021gbh,AbdulKhalek:2022hcn}. 
The EIC, which is expected to start operating in the 2030s, will collide
proton and lepton beams, with the possibility of polarising them both.
Spin asymmetries for longitudinally polarised inclusive DIS and semi-inclusive
DIS (SIDIS) will be measured for a range of proton momentum fractions $x$ and
electroweak boson virtualities $Q^2$ that will significantly extend those
accessed by current measurements (see {\it e.g.} Fig.~(1) in
Ref.~\cite{Ethier:2020way}). 
The precision of these measurements will be unprecedented and is forecast to
attain the percent level (see {\it e.g.}~Sect.~II in Ref.~\cite{Borsa:2020lsz}).

The precision of future EIC measurements calls for a matching accuracy
of the corresponding theoretical predictions. Perturbative corrections
to the massless polarised inclusive DIS structure function $g_1$ have
been known at next-to-next-to-leading order (NNLO) in the
strong-coupling expansion, {\it i.e.}~$\mathcal{O}(\alpha_s^2)$, for a
long time~\cite{Zijlstra:1993sh}\footnote{Other massless
  polarised DIS structure functions have been recently computed up to
  NNLO accuracy as well~\cite{Borsa:2022irn}.}, and up to N$^3$LO, {\it
  i.e.}~$\mathcal{O}(\alpha_s^3)$, since very recently~\cite{Blumlein:2022gpp}.
Massive contributions are known up to
$\mathcal{O}(\alpha_s^2)$~\cite{Hekhorn:2018ywm}, together with their
asymptotic limit~\cite{Behring:2015zaa,Ablinger:2019etw,Behring:2021asx,
  Blumlein:2021xlc,Bierenbaum:2022biv,Ablinger:2023ahe}. In the last two years,
NNLO corrections have also been computed for $W$-boson production in polarised
proton--proton collisions~\cite{Boughezal:2021wjw}, and for the massless
polarised SIDIS structure function $g_1^h$. In this latter case the
computation was performed first using an approximation based on
the threshold resummation formalism~\cite{Abele:2021nyo}, then exactly
using various analytical methods~\cite{Bonino:2024wgg,Goyal:2024tmo,
  Goyal:2024emo}.

The accuracy of theoretical predictions for spin-dependent observables, however,
depends not only on the accuracy of the perturbative computations, but also on
the accuracy of the PDFs that must be convolved with them.
The latter are determined by comparing theoretical predictions
computed with the same accuracy to the experimental data. The first polarised
PDF set accurate to NNLO was determined by
analysing polarised DIS data only~\cite{Taghavi-Shahri:2016idw}. This
data is not sensitive to the decomposition of the proton spin into the
separate contributions carried by quarks and antiquarks of different
flavours. Hence, other polarised PDF
sets~\cite{deFlorian:2014yva,Nocera:2014gqa} based on more global data
sets, despite being accurate only to next-to-leading order (NLO), have
been more widely used so far. Very recently, two new global analyses
of polarised PDFs accurate to NNLO were completed,
{\sc BDSSV}~\cite{Borsa:2024mss} and
{\sc NNPDFpol2.0}~\cite{Cruz-Martinez:2025ahf}.
The {\sc BDSSV} analysis incorporates inclusive and semi-inclusive DIS,
single-inclusive jet, gauge boson, and hadron production
measurements in polarised proton--proton collisions. Whereas theoretical
predictions for inclusive DIS and gauge boson production measurements are
computed exactly at NNLO, the approximations of Refs.~\cite{Abele:2021nyo,
  deFlorian:2007tye,Hinderer:2018nkb} are used to analyse SIDIS data and
single-inclusive jet and hadron production data in proton--proton collisions,
respectively. The {\sc NNPDFpol2.0} analysis incorporates inclusive DIS,
and single-inclusive jet, di-jet, and gauge boson production measurements in
proton--proton collisions. Theory uncertainties, due to missing (unknown)
higher orders in the matrix elements and in the splitting functions, are
systematically taken into account by means of the formalism of
Refs.~\cite{NNPDF:2019vjt,NNPDF:2019ubu,NNPDF:2024dpb}.

In this paper we present {\sc MAPPDFpol1.0}, a global analysis of polarised
PDFs accurate to NNLO based on DIS and SIDIS data. The novelty is the use of
a fitting framework similar to that developed in two previous works in which we
determined the pion and kaon Fragmentation Functions
(FFs)~\cite{Khalek:2021gxf,AbdulKhalek:2022laj}. This framework
combines a neural-network parametrisation of polarised PDFs (optimised
through knowledge of the analytical derivative of neural networks with
respect to their parameters) with a Monte Carlo representation of PDF
uncertainties.  This approach --- which has been extensively used by
the NNPDF Collaboration to determine the unpolarised/polarised proton and
nuclear PDFs and the FFs (see {\it e.g.} Refs.~\cite{NNPDF:2021njg,
  Nocera:2014gqa,Cruz-Martinez:2025ahf,AbdulKhalek:2022fyi,Bertone:2017tyb}
and references therein)
--- allows one to reduce the parametrisation bias as much as possible
and to faithfully propagate experimental uncertainties into PDFs.
These features distinguish this approach from that deployed in
Ref.~\cite{Borsa:2024mss}, where a simpler functional form, possibly subject to
parametrisation bias, is used instead. We make use of the approximate NNLO
corrections to the SIDIS matrix elements~\cite{Abele:2021nyo},
as done in Ref.~\cite{Borsa:2024mss} and in our previous
determination of FFs~\cite{Khalek:2021gxf,AbdulKhalek:2022laj}. However,
by comparing results for theoretical predictions obtained with either
the approximate or exact NNLO computations~\cite{Bonino:2024wgg,Goyal:2024tmo},
we show that our polarised PDFs remain relatively robust.
We carefully assess the impact of perturbative corrections, data sets, and
theoretical constraints on the polarised PDFs and their uncertainties.
In particular, we test possible violations of the SU(2) and SU(3) flavour
symmetries, the effect of positivity constraints, and, for the first time,
whether SIDIS data combined with NNLO corrections allow one to determine an
asymmetry between polarised strange and antistrange PDFs.

The structure of the paper is as follows. In Sect.~\ref{sec:exp} we
present the experimental data analysed in this work. In
Sect.~\ref{sec:theory} we discuss the details of the corresponding
theoretical computations. In Sect.~\ref{sec:methodology} we review the
methodological aspects of the analysis, focusing on those that are
peculiar to the determination of polarised PDFs. The results of our
determination are presented in Sect.~\ref{sec:results}, in which we
discuss in turn the impact of NNLO corrections, of data sets, and of
theoretical constraints. A summary and an outlook are finally given
in Sect.~\ref{sec:conclusions}.  The {\sc MAPPDFpol1.0} sets are
released in the {\sc LHAPDF} format~\cite{Buckley:2014ana}, and the
software developed to produce them is made open
source~\cite{valerio_bertone_2024_10933177}.

\section{Experimental data}
\label{sec:exp}

This analysis is based on a comprehensive set of measurements of
polarised structure functions in lepton-nucleon DIS and
SIDIS. Concerning DIS, we consider data coming from the
EMC~\cite{EuropeanMuon:1989yki}, SMC~\cite{SpinMuon:1998eqa}, and
COMPASS~\cite{COMPASS:2015mhb,COMPASS:2016jwv} experiments at CERN,
from the E142~\cite{E142:1996thl}, E143~\cite{E143:1998hbs},
E154~\cite{E154:1997xfa}, and E155~\cite{E155:2000qdr} experiments at
SLAC, from the HERMES~\cite{HERMES:1997hjr,HERMES:2006jyl} experiment
at DESY, and from the Hall A~\cite{JeffersonLabHallA:2016neg} and
CLAS~\cite{CLAS:2014qtg} experiments at JLab. All of these experiments
deliver data for the polarised inclusive structure function $g_1$,
reconstructed from the longitudinal double spin asymmetry (see
Sect.~2.1 in Ref.~\cite{Ball:2013lla} for details), except E155, Hall
A, and CLAS, which instead deliver data for $g_1$ normalised to the
unpolarised inclusive structure function $F_1$.  Concerning SIDIS, we
consider data from COMPASS~\cite{COMPASS:2010hwr} and
HERMES~\cite{HERMES:2018awh}. Both of these experiments deliver data
for the polarised semi-inclusive structure function $g_1^h$, with
$h=\pi^+,\pi^-,K^+,K^-$, normalised to its unpolarised counterpart
$F_1^h$. The target in the aforementioned DIS and SIDIS data sets is
a proton, or a neutron, or a deuteron.

\begin{figure}[!t]
  \centering
  \includegraphics[width=0.85\textwidth]{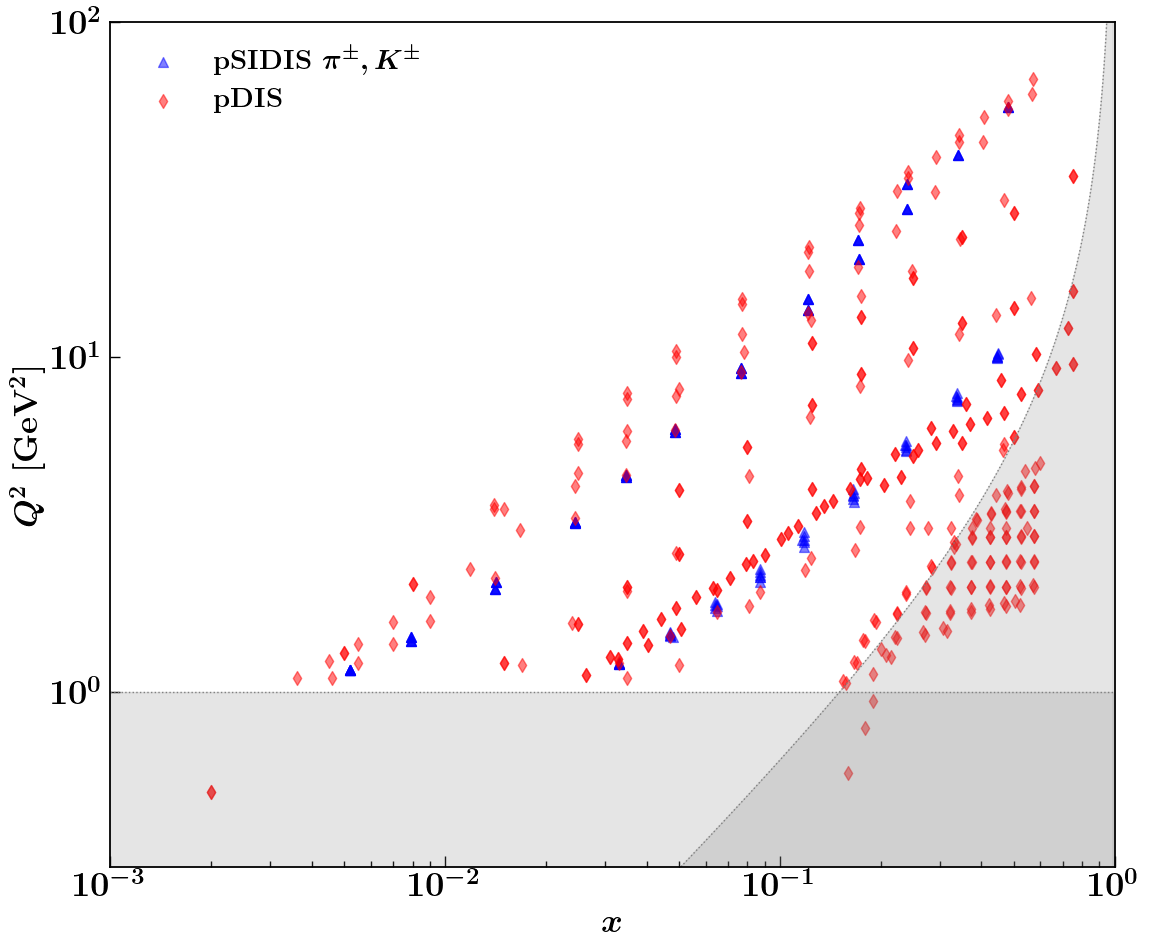}
  \caption{\small Kinematic coverage in the $(x,Q^2)$ plane of the
    polarised DIS and SIDIS data included in this analysis. The shaded
    areas correspond to regions excluded by the cuts in $Q^2$ and
    $W^2$, see text for additional details.}
  \label{fig:data}
\end{figure}

The data set is summarised in Table~\ref{tab:chi2} and its kinematic
coverage in the ($x,Q^2$) plane is displayed in
Fig.~\ref{fig:data}. The data points cover a rather limited region,
roughly $0.005\lesssim x\lesssim 0.5$ and $1\lesssim Q^2\lesssim 100$~GeV$^2$.
We apply kinematic cuts on the virtuality $Q^2\geq Q^2_{\rm cut}$ and on
the invariant mass of the final state $W^2=Q^2(1-x)/x\geq W^2_{\rm cut}$ as
default. For SIDIS data, we consider an optional, additional cut
$x\geq x_{\rm cut}$, where $x$ is the initial proton momentum fraction carried by
the struck parton. The cut on $Q^2$ removes a region where perturbative
computations become unreliable because of the rise of the strong
coupling. The cut on $W^2$ removes a region where leading-twist
factorisation, on which we base our theoretical framework, becomes
likewise insufficient. The cut on $x$ removes a region where the
approximate computation of NNLO SIDIS matrix elements~\cite{Abele:2021nyo},
which we use in our analysis, differ from the exact
computation~\cite{Bonino:2024wgg,Goyal:2024tmo}.
We choose $Q^2_{\rm cut}=1$~GeV$^2$, $W^2=6.25$~GeV$^2$,
and $x_{\rm cut}=0.1$. The first value is a common choice in several
determinations of unpolarised and polarised PDFs, and results from a
tradeoff between incorporating as much experimental information as
possible and preserving the reliability of perturbative
computations. The second value is selected among the set of values
$W^2_{\rm cut}=1.0,4.0,6.25,9.0$~GeV$^2$ after performing a fit with each
of them and verifying that it maximises the fit quality and preserves
the stability of PDFs in comparison to those obtained with other choices.
The third value is chosen so that SIDIS asymmetries computed with the
approximate or exact computations differ by an amount smaller than the
experimental uncertainty.
The regions excluded by the first and second kinematic cuts correspond to the
shaded areas in Fig.~\ref{fig:data}. Note that some JLab measurements, namely
those from Hall-A~\cite{JeffersonLabHallA:2004tea} and CLAS~\cite{CLAS:2015otq},
are completely excluded by our kinematic cuts.  They will therefore not be
considered in the rest of this paper.

Statistical and systematic uncertainties are separately provided for each of
the measurements listed in Table~\ref{tab:chi2}, however detailed
information on correlations is missing in most cases. Specifically, for the
EMC, E143, and E155 experiments a correlated multiplicative uncertainty is
specified, whereas for the HERMES DIS and SIDIS experiments of
Refs.~\cite{HERMES:2006jyl,HERMES:2018awh} the bin-by-bin covariance matrix
is given. In all of these cases we take into account the available piece of
information on experimental correlations. In the other cases, we
assume the systematic uncertainties to be uncorrelated and we add them in
quadrature with the statistical ones.

In addition to the aforementioned DIS and SIDIS measurements, we also
optionally consider data corresponding to semi-leptonic $\beta$-decays
of the baryonic octet. Assuming SU(2) and SU(3) flavour symmetries,
these can be related (see {\it e.g.} Ref.~\cite{Ratcliffe:2004jt}) to
the lowest moments of the triplet and octet polarised PDF combinations
defined as
\begin{equation}
  a_3
  =
  \int_{0}^{1} dx \,
  \left[ \Delta f_u^{+}(x,Q^2) - \Delta f_d^{+}(x,Q^2) \right] \,,
  \label{eq:a3_PM}
\end{equation}
\begin{equation}
  a_8
  =
  \int_{0}^{1} dx \,
  \left[ \Delta f_u^{+}(x,Q^2) + \Delta f_d^{+}(x,Q^2)
    - 2 \Delta f_s^{+}(x,Q^2) \right] \,,
  \label{eq:a8_PM}
\end{equation}
where $\Delta f_q^{+}=\Delta f_q+\Delta f_{\bar{q}}$. Both $a_3$ and
$a_8$ are scale independent, however, for the practical purpose of
computing Eqs.~\eqref{eq:a3_PM}-\eqref{eq:a8_PM}, we set
$Q^2=1$~GeV$^2$. The values that we use are $a_3=1.2756\pm 0.0013$ and
$a_8=0.585 \pm 0.025$~\cite{ParticleDataGroup:2022pth}. The impact of
these data points will be discussed in Sect.~\ref{subsec:theory}.

\section{Theoretical predictions}
\label{sec:theory}

The experimental data described in Sect.~\ref{sec:exp} are measured in
DIS and SIDIS processes where a polarised lepton beam scatters off a
polarised nucleon target
\begin{equation}
  \ell(k,s) + N(P, S) \longrightarrow \ell'(k') + (h^{\pm} (p_h)) + X \,.
  \label{eq:DIS}
\end{equation}
Here $P$ is the four-momentum of the nucleon $N$, $k$ ($k'$) is the
four-momentum of the incoming (outgoing) lepton $\ell$ ($\ell'$), and
$p_h$ is the four-momentum of the outgoing hadron $h^\pm$ (in SIDIS);
$S$ and $s$ are the spin four-vectors of the nucleon and incoming lepton,
respectively. Because of the very moderate values of the virtuality $Q^2$
(see Fig.~\ref{fig:data}), the scattering involves only the exchange of a
virtual photon.

The measured observables are related to the difference between cross sections
with opposite target spins, which, neglecting terms suppressed by powers of
$M^2/Q^2$ with $M$ the nucleon mass, read
\begin{equation}
  \begin{array}{lcl}
    \textrm{DIS}
    \hspace{-25mm}
    &:&
    \displaystyle
    \hspace{3mm}
    \frac{d\Delta\sigma}{dxdy}
    =
    \frac{1}{2}\left(\frac{d\sigma^{\rightarrow,\Rightarrow}}{dxdy}
    -
    \frac{d\sigma^{\rightarrow,\Leftarrow}}{dxdy}\right)
    =
    \frac{4\pi\alpha^2}{xQ^2} (2-y) g_1(x,Q^2) \,,
    \\
    \\
    \textrm{SIDIS}
    \hspace{-25mm}
    &:&
    \displaystyle
    \hspace{3mm}
    \frac{d\Delta\sigma_h}{dxdy}
    =
    \frac{1}{2}\left(\frac{d\sigma_h^{\rightarrow,\Rightarrow}}{dxdy}
    -
    \frac{d\sigma_h^{\rightarrow,\Leftarrow}}{dxdy}\right)
    =
    \frac{4\pi\alpha^2}{xQ^2} (2 - y) g_1^{h}(x,z,Q^2) \,.
  \end{array}
  \label{eq:cs_diff_l}
\end{equation}
Here $\rightarrow$ denotes the longitudinal polarisation of the
incoming lepton, parallel to its four-momentum, while $\Rightarrow$
($\Leftarrow$) denotes the longitudinal polarisations of the nucleon
parallel (antiparallel) to the lepton four-momentum. The variables
appearing in Eq.~\eqref{eq:cs_diff_l} are Lorentz invariant and are
defined as follows: $Q^2=-q^2$ is the (negative) virtuality of the
photon; $x=Q^2/(2P\cdot q)$ is the (lowest) momentum fraction of the
initial-state nucleon carried by the scattering parton;
$z=P\cdot p_h/(P\cdot q)$ is the (lowest) momentum fraction of the
fragmenting parton carried by the identified final-state hadron; and
$y=P\cdot q/(P\cdot k)$ is the inelasticity, that is the energy
fraction transferred by the incoming lepton.

The rightmost equalities in Eq.~\eqref{eq:cs_diff_l}, where $\alpha$
is the fine-structure constant, define the polarised structure
functions $g_1$ and $g_1^{h}$, which are reconstructed from the cross
section differences measured by the experimental collaborations. Using
the leading-twist collinear factorisation theorem valid for
$Q^2 \gg \Lambda_{\textrm{QCD}}$, these structure functions factorise
as\footnote{Note that the factorised expression for the SIDIS
  structure function $g_1^{h}$ does not include a
  $\Delta\mathcal{C}_{gg}$ term proportional to both the gluon PDF
  $\Delta f_g$ and the gluon fragmentation function $D_g^h$ that is in
  principle present starting from NNLO. However, as discussed below,
  in our analysis we use an approximation for the NNLO corrections to
  SIDIS that does not include this term.}
\begin{align}
  &g_1 (x,Q^2)
  =
  \frac{1}{2} x \sum_{q} e_{q}^2
  \left\{
  \Delta f_q (x,Q^2)  \mathop{\otimes}_{x}  \Delta \mathcal{C}_{q}(x,Q^2)
  +
  \Delta f_g (x,Q^2) \mathop{\otimes}_{x}  \Delta \mathcal{C}_{g}(x,Q^2)
  \right\} \,,
  \label{eq:g1_DIS}
  \\[5pt]
  &g_1^{h} (x,z,Q^2)
  =
  \frac{1}{2} x \sum_{q} e_{q}^2
  \left\{ \left[
    \Delta f_q(x,Q^2) \mathop{\otimes}_{x} \Delta \mathcal{C}_{qq}(x,z,Q^2)
    + \Delta f_g(x,Q^2) \mathop{\otimes}_{x} \Delta \mathcal{C}_{qg}(x,z,Q^2)
    \right]
  \mathop{\otimes}_{z} D_{q}^{h}(z,Q^2)
  \right.
  \notag\\
  & \hspace{42mm}
  \left.
  +\Delta f_q(x,Q^2) \mathop{\otimes}_{x} \Delta \mathcal{C}_{gq}
  \mathop{\otimes}_{z} D^{h}_{g}(z,Q^2)
  \right\} \,,
  \label{eq:g1_SIDIS}
\end{align}
where $\displaystyle \mathop{\otimes}_{w}$ denotes the usual Mellin
convolution acting on the variable $w$ as follows:
\begin{equation}
  C(w) \mathop{\otimes}_{w} h(w)
  =
  \int_{w}^{1} \frac{dw'}{w'} C(w') h\left(\frac{w'}{w} \right) \,.
\end{equation}
In Eqs.~\eqref{eq:g1_DIS} and~\eqref{eq:g1_SIDIS} the sum runs over
the quark and antiquark flavours active at scale $Q^2$, $e_q$ is the
electric charge of the flavour $q$, $\Delta \mathcal{C}$ are the
appropriate coefficient functions, and $\Delta f_{q(g)}$ and
$D^h_{q(g)}$ are the longitudinally polarised quark (gluon) PDF of the
proton and the unpolarised quark (gluon) FF of the hadron $h$,
respectively.  Since some experiments deliver data for the ratios
$g_1/F_1$ and $g_1^{h}/F_1^{h}$ (see Table~\ref{tab:chi2}), the
unpolarised DIS and SIDIS structure functions $F_1$ and $F_1^{h}$ have
to be computed as well. This is done by replacing the polarised PDFs
and coefficient functions $\Delta f_i$ and $\Delta\mathcal{C}$ in
Eqs.~\eqref{eq:g1_DIS} and~\eqref{eq:g1_SIDIS} with their unpolarised
counterparts $f_i$ and $\mathcal{C}$. We finally assume exact isospin
symmetry to relate neutron and proton structure functions, which in
turn also allows us to reconstruct the structure functions of the
deuteron. No nuclear corrections are taken into account in the case of
a deuterium target. Likewise, we do not consider target mass corections.

In Eqs.~\eqref{eq:g1_DIS} and~\eqref{eq:g1_SIDIS} the coefficient
functions $\Delta \mathcal{C}$ are computed as a perturbative series
in the strong coupling $\alpha_s$ 
\begin{equation}
  \Delta \mathcal{C} (k,Q)
  =
  \sum_{n=0} \left( \frac{\alpha_s(Q^2)}{4\pi} \right)^n
  \Delta \mathcal{C}^{(n)} (k)\,,
  \label{eq:coeff_func}
\end{equation}
where the kinematic variable $k$ corresponds to the initial-state momentum
fraction $x$ for DIS and to the pair of initial- and final-state momentum
fractions $(x,z)$ for SIDIS. We consider massless-quark coefficient functions
up to NNLO. These were computed up to NNLO in Ref.~\cite{Zijlstra:1993sh} for
DIS and up to NLO in Refs.~\cite{Furmanski:1981cw, deFlorian:1997zj} for SIDIS.
We consider approximate NNLO coefficient functions for SIDIS, as derived in
Ref.~\cite{Abele:2021nyo} by extracting the fixed-order dominant
contributions associated to the emission of soft gluons close to
production threshold. This approximation produces corrections only to
the $\Delta\mathcal{C}_{qq}$ coefficient function. The reliability of this
approximation was recently checked against the exact computation in the
unpolarised~\cite{Goyal:2023xfi,Bonino:2024qbh} and polarised
cases~\cite{Bonino:2024wgg,Goyal:2024tmo}, finding excellent agreement for
sufficiently large values of $x$ and $z$. In Sect.~\ref{sec:results}
we will compare theoretical predictions obtained with the approximate and
exact computations, with fixed PDFs, finding good agreement between the two
in the kinematic region covered by the experimental data.

We do not consider massive-quark corrections, which, albeit being known for
$g_1$ up to $\mathcal{O}(\alpha_s^2)$~\cite{Hekhorn:2018ywm,Behring:2015zaa,
  Ablinger:2019etw,Behring:2021asx,Blumlein:2021xlc,Bierenbaum:2022biv,
  Ablinger:2023ahe}, are likely to be small in comparison to the size of the
experimental uncertainties of the data included in the
fit~\cite{Hekhorn:2024tqm}. Intrinsic heavy-quark distributions are assumed to
be identically zero below the corresponding thresholds. At higher scales,
heavy-quark distributions are perturbatively generated by means of DGLAP
evolution in the zero-mass variable-flavour-number scheme
(ZM-VFNS).\footnote{We note that evolution
  in the VFNS requires matching conditions that encode
  the perturbative transition between adjacent schemes differing by
  one unit in the number of active quark flavours. NNLO evolution
  would imply the use of matching conditions accurate to
  $\mathcal{O}(\alpha_s^2)$. Despite the full set of
  $\mathcal{O}(\alpha_s^2)$ corrections to the polarised matching
  conditions has been recently computed in
  Ref.~\cite{Bierenbaum:2022biv}, they are not presented in a format
  readily implementable in our framework. Specifically, the
  expressions are presented in Mellin space, rather than in momentum
  space, and for heavy-quark masses renormalised in the
  $\overline{\mbox{MS}}$ scheme, rather than in the pole-mass
  scheme. Therefore, we use $\mathcal{O}(\alpha_s)$ matching
  conditions also when evolving polarised PDFs at NNLO, leaving the
  implementation of the $\mathcal{O}(\alpha_s^2)$ corrections for a
  future work.}
Perturbative corrections to the splitting functions entering the DGLAP
equations are taken into account consistently up to
NNLO accuracy~\cite{Moch:2014sna,Moch:2015usa,Blumlein:2021ryt}.

All the expressions for the aforementioned unpolarised and polarised coefficient
and splitting functions are implemented in the public
code {\sc APFEL++}~\cite{Bertone:2013vaa,Bertone:2017gds}, which
we use to compute the theoretical predictions that enter the fit.
The values of the relevant physical parameters are as follows:
$\alpha_s(M_Z)=0.118$, $m_c=1.51$~GeV, and $m_b=4.92$~GeV.

\section{Fitting methodology}
\label{sec:methodology}

The methodological framework used to infer polarised PDFs from data is the same
used to determine FFs in Refs.~\cite{Khalek:2021gxf,AbdulKhalek:2022laj}.
Its main ingredients are the propagation of experimental
uncertainties into polarised PDFs by means of Monte Carlo sampling, the
parametrisation of PDFs by means of neural networks, and the optimisation of
the parameters by means of analytic gradient descent minimisation.

Concerning the Monte Carlo sampling, all of our polarised PDF sets are made of
$N_{\rm rep}=150$ Monte Carlo replicas. This number was chosen by requiring that
the statistical features of the experimental data, namely central values,
uncertainties, and correlations, be reproduced by averages, standard
deviations, and covariances computed over the Monte Carlo ensemble with an
accuracy below 1\%.

Concerning the parametrisation, we utilise a single one-layered
feed-forward neural network with one input node corresponding to the momentum
fraction $x$, 10 intermediate nodes, and 7 output nodes, all with a sigmoid
activation function. This architecture amounts to a total of 97 free
parameters and was selected (among several with as few as 5 nodes and as many
as 20 nodes in the hidden layer) as the one corresponding to a sufficiently
flexible parametrisation to accommodate all data points without increasing
its complexity (and thus the computational burden to train it) too much.
We note that this architecture is simpler than the one used in our earlier
works on FFs~\cite{Khalek:2021gxf,AbdulKhalek:2022laj} because of the
comparatively scarcer and less precise measurements included in the fit.

The output nodes correspond to the independent polarised PDFs that we fit,
namely
\begin{equation}
  \left\{
  \Delta f_u,\,\Delta f_{\bar{u}},\,\Delta f_d,
  \,\Delta f_{\bar{d}},\,\Delta f_s,\,\Delta f_{\bar{s}},
  \,\Delta f_g
  \right\}\,.
  \label{eq:basis}
\end{equation}
Note that, for the first time, we allow $\Delta f_s$ to be different from
$\Delta f_{\bar{s}}$. The availability of SIDIS data for production of positively
and negatively charged kaons and of NNLO corrections, which make polarised
strange quarks and antiquarks evolve differently, can in principle distinguish
between the two distributions. The parametrisation scale is set to 
$Q_0^2=1$~GeV$^2$.

The output layers, for each Monte Carlo replica $k$, are shifted and rescaled as
\begin{equation}
  \Delta f_{i}^{(k)} (x, Q_0^2)
  =
  \left[ 2\,{\rm NN}_i(x) - 1 \right]
  f_{i}^{(U(1,100))} (x,Q_0^2)\,,
  \quad
  i=g,u,\bar{u},d,\bar{d},s,\bar{s}\,,
  \label{eq:pos_net}
\end{equation}
where $U(1,100)$ is a random integer number uniformly sampled in the
interval $[1,100]$, which denotes a replica in a given unpolarised PDF
set.  In this way, each output node, corresponding to a replica $k$ of
each polarised PDF in the basis of Eq.~\eqref{eq:basis}, is bound by
an unpolarised PDF replica
\begin{equation}
  \left| \Delta f_i^{(k)} (x,Q_0^2) \right|
  \leq
  f_i^{(U(1,100))} (x,Q_0^2)\,, \quad \forall x\,.
  \label{eq:positivity_random}
\end{equation}
The parametrisation in Eq.~\eqref{eq:pos_net} enforces by construction the
constraint
\begin{equation}
  \left| \Delta f_i (x,Q^2) \right| \leq f_i (x,Q^2)\,,\quad \forall x\,, \forall Q^2\,,
  \label{eq:positivity_fl}
\end{equation}
which follows, at leading order (LO), from requiring positivity of cross
sections
\begin{equation}
  \begin{split}
    \left| g_1 (x,Q^2) \right|& \leq F_1(x, Q^2) \,,\\
    \left| g_1^{h} (x,z,Q^2)  \right|& \leq F_1^{h}(x, z, Q^2) \,.
  \end{split}
  \label{eq:positivity_obs}
\end{equation}
Beyond LO, more complicated conditions hold~\cite{Altarelli:1998gn,
  deFlorian:2024utd}, however they differ only mildly from
Eq.~\eqref{eq:positivity_fl} at large $x$ where they are relevant to
constrain PDFs (see {\it e.g.}\ Fig.~3 in~\cite{Altarelli:1998gn}).
The uncertainty on the bound in Eq.~\eqref{eq:positivity_fl} due to
the uncertainty of the unpolarised PDF is incorporated into the
polarised PDF thanks to the fact that an unpolarised PDF replica is
chosen at random for each polarised PDF replica in
Eq.~\eqref{eq:positivity_random}.  The parametrisation in
Eq.~\eqref{eq:pos_net} has two additional advantages: first, it
guarantees that polarised PDFs vanish at $x=1$ as a consequence of the
fact that unpolarised PDFs also do so; second, it guarantees that
polarised PDFs are integrable over $x$, as required to ensure the
finiteness of their moments.

Whereas Eq.~\eqref{eq:pos_net} is our default parametrisation, we have also
tried the alternative implementation
\begin{equation}
  \Delta f_{i}^{(k)} (x, Q_0^2)
  = \left[ 2\,{\rm NN}_i(x) - 1 \right]
  \left[ f_i^{(0)} (x,Q_0^2)+K\sigma_i (x,Q_0^2)\right]\,,
  \quad i=g,u,\bar{u},d,\bar{d},s,\bar{s} \,,
  \label{eq:pos_net_alt}
\end{equation}
where $f_i^{(0)}$ is the unpolarised PDF central replica, $\sigma_i$ is the
one-sigma unpolarised PDF uncertainty, and $K$ is a positive integer. The
larger the value of $K$, the looser the positivity constraint. The
implementation of Eq.~\eqref{eq:pos_net_alt} allowed us to study how
polarised PDFs depend on the positivity constraint,
see Sect.~\ref{subsec:theory}.

The unpolarised PDFs entering Eqs.~\eqref{eq:pos_net}
and~\eqref{eq:pos_net_alt} are taken from the {\sc
  NNPDF3.1}~\cite{NNPDF:2017mvq} parton set with perturbative
charm. This parton set is also used to compute the unpolarised
structure functions $F_1$ ($F_1^h$) when the DIS (SIDIS) data is
presented as $g_1/F_1$ ($g_1^h/F_1^h$). In the SIDIS case, a set of
FFs is also needed, which we take from the {\sc
  MAPFF1.0}~\cite{Khalek:2021gxf,AbdulKhalek:2022laj} set.\footnote{We
  use the {\sc NNPDF3.1} PDF set, despite the availability of the more
  recent {\sc NNPDF4.0} PDF set~\cite{NNPDF:2021njg}, to maximise
  consistency with {\sc MAPFF1.0}, which used the {\sc NNPDF3.1} PDF
  set as input.} The {\sc NNPDF3.1} and {\sc MAPFF1.0} sets are
respectively obtained from a global analysis of measurements in DIS
and a variety of processes in proton--proton collisions, and from a
global analysis of single-inclusive hadron production in
electron-positron annihilation and SIDIS. Both these sets are based on
a similar methodology and are accurate up to NNLO. The perturbative
accuracy of the PDF and FF sets is chosen consistently with the
accuracy of the analysis of polarised PDFs presented here (NLO or
NNLO). In all cases, PDF and FF replicas are chosen randomly from the
corresponding sets for each fitted polarised PDF replica.

Concerning parameter optimisation, we use the exact same strategy
as in Refs.~\cite{Khalek:2021gxf,AbdulKhalek:2022laj}. Specifically,
for each replica: we minimise the $\chi^2$ (see {\it e.g.}\ Eq.~(21)
in Ref.~\cite{Khalek:2021gxf} for its definition); we adopt
cross-validation with a training fraction of 80\% for data sets with
more than 10 points and 100\% otherwise; we determine the optimal
parameters with the Levenberg-Marquardt algorithm as implemented in
{\sc Ceres-Solver}~\cite{Agarwal_Ceres_Solver_2022}, computing the
gradient of the neural network with respect to the free parameters analytically
using the {\sc NNAD} library~\cite{AbdulKhalek:2020uza}; and we
discard replicas which, at the end of the training, result in a value
of the $\chi^2$ per data point larger than three.\footnote{These outlieres
typically occur once or twice every hundred replicas.}

\section{Results and discussion}
\label{sec:results}

We now present {\sc MAPPDFpol1.0}, the polarised PDF determination
obtained from the experimental data, the theoretical predictions, and
the fitting methodology described in
Sects.~\ref{sec:exp}-\ref{sec:methodology}.  We discuss in turn the
impact of NNLO corrections, data, and theoretical constraints by
comparing fits obtained upon variations of each of these aspects to
the same baseline fit. This baseline fit is obtained using NNLO theoretical
predictions, the positivity constraint as in Eq.~\eqref{eq:pos_net},
and the global data set, including the measurements for $a_3$ and $a_8$.

\subsection{Impact of NNLO corrections}
\label{subsec:nnlo}

Table~\ref{tab:chi2} reports the value of the $\chi^2$ per data point
of the NLO and NNLO {\sc MAPPDFpol1.0} determinations. The fit quality
is generally good for both the individual and the global data sets and
for both the NLO and NNLO fits. However, we remark that the global $\chi^2$
per data point increases from 0.64 to 0.78 when moving from NLO to NNLO.
This difference corresponds to about two standard deviations of the $\chi^2$
per data point distribution. This increase can be seen in both DIS and SIDIS
data, although it is more pronounced for the latter. A similar behaviour was
observed also in the case of FFs when fitting single-inclusive annihilation and
SIDIS data~\cite{AbdulKhalek:2022laj,Borsa:2022vvp}. This observation questions
whether the inclusion of (approximate) perturbative corrections lead to a
better PDF determination or not. On the other hand, we also remark that the
global $\chi^2$ per data point at both perturbative orders is significantly
smaller than the expectation, \textit{i.e.} one. At NLO, by about five standard
deviations of the $\chi^2$ per data point distribution; at NNLO, by about three.
Being the $\chi^2$ per data point of the NNLO fit closer to one, one
should conclude that this is better in statistical terms. The smallness of the
global $\chi^2$ was already observed in previous NLO
analyses~\cite{Ball:2013lla,Nocera:2014gqa,DeFlorian:2019xxt,Ethier:2017zbq},
and is ascribed to a limited knowledge of experimental correlations, which
result in an effective uncertainty overestimate. This observation questions
whether, given the intrinsic limitations of the data set, one should judge the
quality of our PDFs  based only on the value of the $\chi^2$ per data point.
In this respect, we believe that one must be careful, and instead validate the
stability of the PDFs according to additional investigations, that in particular
clarify the interplay between higher-order corrections and the data.
We will present these investigations in Sect.~\ref{subsec:data}.

\begin{table}[!t]
  \centering 
  \footnotesize
  \renewcommand{\arraystretch}{1.4}
  \begin{tabularx}{\textwidth}{Xccrccccc}
  \toprule
  & & & & \multicolumn{2}{c}{baseline $\chi^2/N_{\rm dat}$}
  & \multicolumn{2}{c}{no $a_3$, $a_8$ $\chi^2/N_{\rm dat}$}
  & no pos. $\chi^2/N_{\rm dat}$ \\
  Experiment & Ref. & Observable & $N_{\rm dat}$
  & NLO & NNLO & NLO & NNLO & NNLO \\
  \midrule
  EMC
  & \cite{EuropeanMuon:1989yki} & $g_1^p$
  &  10 & 0.57 & 0.49 & 0.59 & 0.55 & 0.53 \\
  SMC
  & \cite{SpinMuon:1998eqa} & $g_1^p$
  &  12 & 0.29 & 0.32 & 0.31 & 0.29 & 0.29 \\
  & \cite{SpinMuon:1998eqa} & $g_1^d$
  &  12 & 1.34 & 1.20 & 1.35 & 1.19 & 1.27 \\
  E142
  & \cite{E142:1996thl} & $g_1^n$
  &   7 & 0.58 & 0.85 & 0.58 & 0.77 & 0.62 \\ 
  E143
  & \cite{E143:1998hbs} & $g_1^p$
  &  25 & 0.74 & 0.69 & 0.74 & 0.68 & 0.54 \\
  & \cite{E143:1998hbs} & $g_1^d$
  &  25 & 1.30 & 1.23 & 1.28 & 1.25 & 1.32 \\
  E154
  & \cite{E154:1997xfa} & $g_1^n$
  &  11 & 0.22 & 0.20 & 0.23 & 0.22 & 0.31 \\
  E155
  & \cite{E155:2000qdr} & $g_1^p/F_1^p$
  &  22 & 0.66 & 0.85 & 0.66 & 0.79 & 0.93 \\
  & \cite{E155:2000qdr} & $g_1^d/F_1^d$
  &  22 & 0.71 & 0.81 & 0.71 & 0.82 & 0.90 \\
  COMPASS
  & \cite{COMPASS:2016jwv} & $g_1^p$
  &  17 & 0.58 & 0.95 & 0.52 & 0.73 & 0.61 \\
  & \cite{COMPASS:2015mhb} & $g_1^d$
  &  15 & 0.36 & 1.02 & 0.31 & 0.84 & 0.62 \\
  HERMES
  & \cite{HERMES:1997hjr} & $g_1^n$
  &   8 & 0.22 & 0.27 & 0.22 & 0.25 & 0.20 \\
  & \cite{HERMES:2006jyl} & $g_1^p$
  &  14 & 0.46 & 0.53 & 0.49 & 0.52 & 0.50 \\
  & \cite{HERMES:2006jyl} & $g_1^d$
  &  14 & 0.63 & 0.74 & 0.68 & 0.71 & 0.77 \\
  JLAB-E06
  & \cite{JeffersonLabHallA:2016neg} & $g_1^n/F_1^n$
  &   1 & 0.72 & 0.86 & 0.69 & 0.82 & 0.51 \\
  JLAB-EG1
  & \cite{CLAS:2014qtg} & $g_1^p/F_1^p$
  &   2 & 0.01 & 0.01 & 0.01 & 0.01 & 0.01 \\
  & \cite{CLAS:2014qtg} & $g_1^d/F_1^d$
  &   1 & 0.01 & 0.01 & 0.01 & 0.01 & 0.01 \\
  \midrule
  Total DIS  & &
  & 218 & 0.48 & 0.55 & 0.48 & 0.53 & 0.52 \\
  \midrule
  COMPASS
  & \cite{COMPASS:2010hwr} & $g_1^{p,\pi^+}/F_1^{p,\pi^+}$
  &  12 & 2.32 & 2.01 & 2.33 & 1.94 & 1.38 \\
  & \cite{COMPASS:2010hwr} & $g_1^{p,\pi^-}/F_1^{p,\pi^-}$
  &  12 & 1.34 & 1.13 & 1.29 & 1.09 & 0.91 \\
  & \cite{COMPASS:2010hwr} & $g_1^{p,K^+}/F_1^{p,K^+}$
  &  12 & 0.69 & 0.94 & 0.77 & 0.97 & 0.82 \\
  & \cite{COMPASS:2010hwr} & $g_1^{p,K^-}/F_1^{p,K^-}$
  &  12 & 0.73 & 0.98 & 0.65 & 0.88 & 0.92 \\
  & \cite{COMPASS:2010hwr} & $g_1^{d,\pi^+}/F_1^{d,\pi^+}$
  &  10 & 0.31 & 1.23 & 0.31 & 1.38 & 0.50 \\
  & \cite{COMPASS:2010hwr} & $g_1^{d,\pi^-}/F_1^{d,\pi^-}$
  &  10 & 0.47 & 1.51 & 0.49 & 1.50 & 0.63 \\
  & \cite{COMPASS:2010hwr} & $g_1^{d,K^+}/F_1^{d,K^+}$
  &  10 & 0.40 & 0.40 & 0.28 & 0.24 & 0.34 \\
  & \cite{COMPASS:2010hwr} & $g_1^{d,K^-}/F_1^{d,K^-}$
  &  10 & 0.92 & 0.82 & 0.88 & 0.70 & 0.79 \\
  HERMES
  & \cite{HERMES:2018awh} & $g_1^{p,\pi^+}/F_1^{p,\pi^+}$
  &   9 & 1.90 & 2.05 & 1.87 & 1.79 & 1.88 \\
  & \cite{HERMES:2018awh} & $g_1^{p,\pi^-}/F_1^{p,\pi^-}$
  &   9 & 1.03 & 0.68 & 1.00 & 0.65 & 0.72 \\
  & \cite{HERMES:2018awh} & $g_1^{d,\pi^+}/F_1^{d,\pi^+}$
  &   9 & 0.35 & 1.53 & 0.34 & 1.41 & 0.82 \\
  & \cite{HERMES:2018awh} & $g_1^{d,\pi^-}/F_1^{d,\pi^-}$
  &   9 & 1.30 & 2.41 & 1.37 & 2.16 & 1.71 \\
  & \cite{HERMES:2018awh} & $g_1^{d,K^+}/F_1^{d,K^+}$
  &   9 & 1.48 & 1.72 & 1.46 & 1.65 & 1.89 \\
  & \cite{HERMES:2018awh} & $g_1^{d,K^-}/F_1^{d,K^-}$
  &   9 & 0.63 & 1.04 & 0.64 & 1.06 & 0.65 \\
  \midrule
  Total SIDIS & &
  & 142 & 1.00 & 1.31 & 0.99 & 1.23 & 0.99\\
  \midrule
  Total      & &
  & 362 & 0.64 & 0.78 & 0.63 & 0.74 & 0.66 \\
  \bottomrule
\end{tabularx}
\\
  \vspace{0.5cm}
  \caption{The data sets, number of data points, and $\chi^2$ per data
    point for the {\sc MAPPDFpol1.0} PDF sets. We display the values
    obtained from the baseline NLO and NNLO fits, from the NLO and
    NNLO fit variants without data for $a_3$ and $a_8$, and from the
    NNLO fit variant with a very loose positivity constraint, see
    Sect.~\ref{subsec:theory}. The $\chi^2$ values are displayed for
    the individual and global data sets.}
  \label{tab:chi2}
\end{table}

\begin{figure}[!t]
  \centering
  \includegraphics[width=0.48\textwidth]{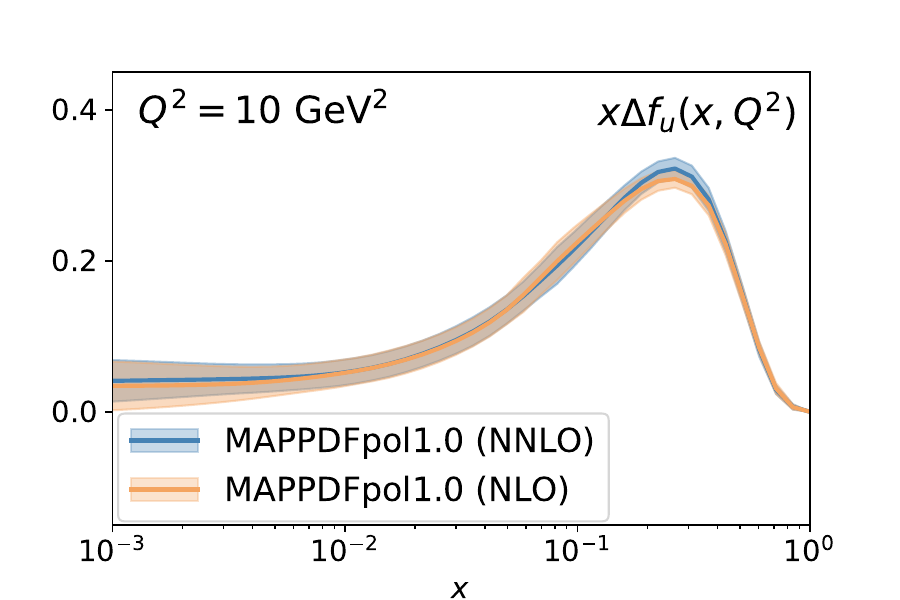}
  \includegraphics[width=0.48\textwidth]{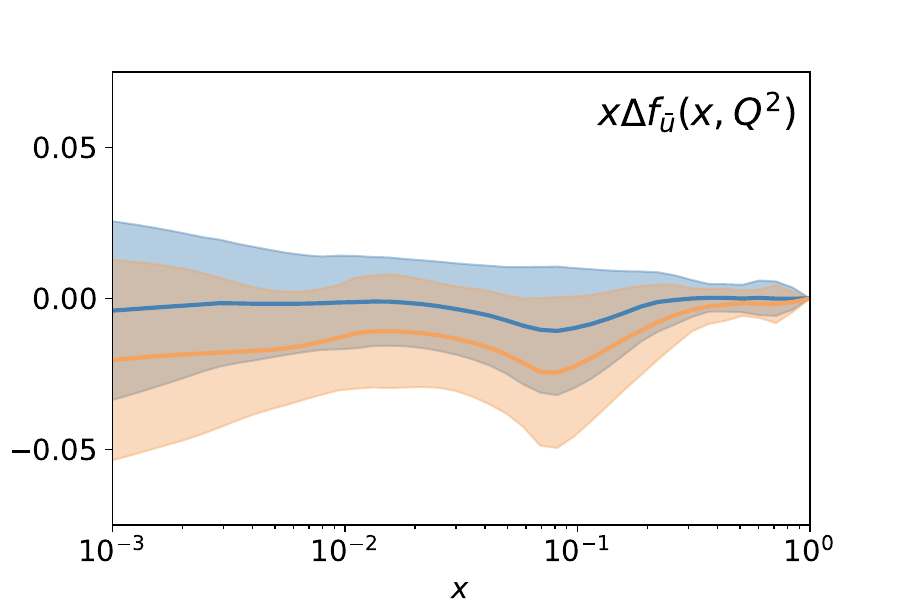}\\
  \includegraphics[width=0.48\textwidth]{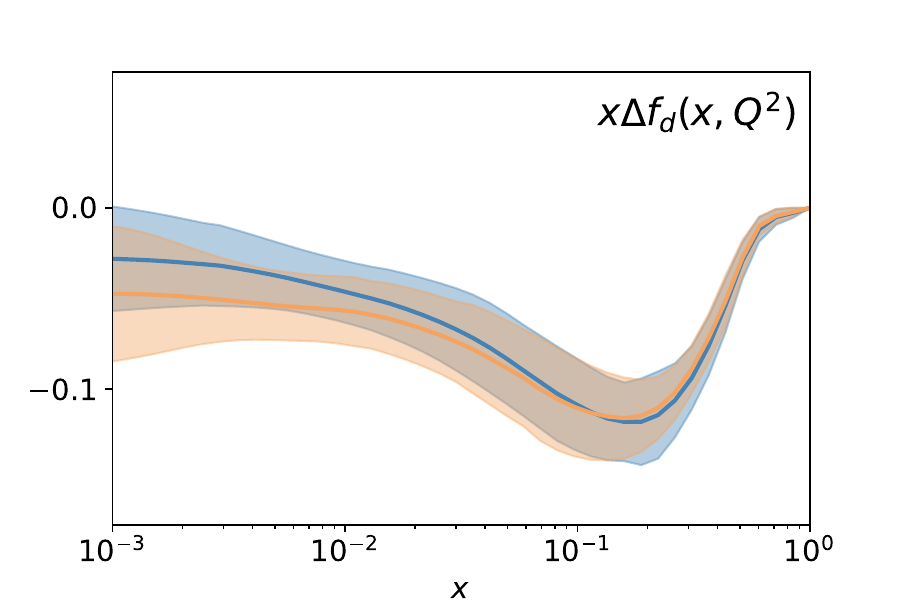}
  \includegraphics[width=0.48\textwidth]{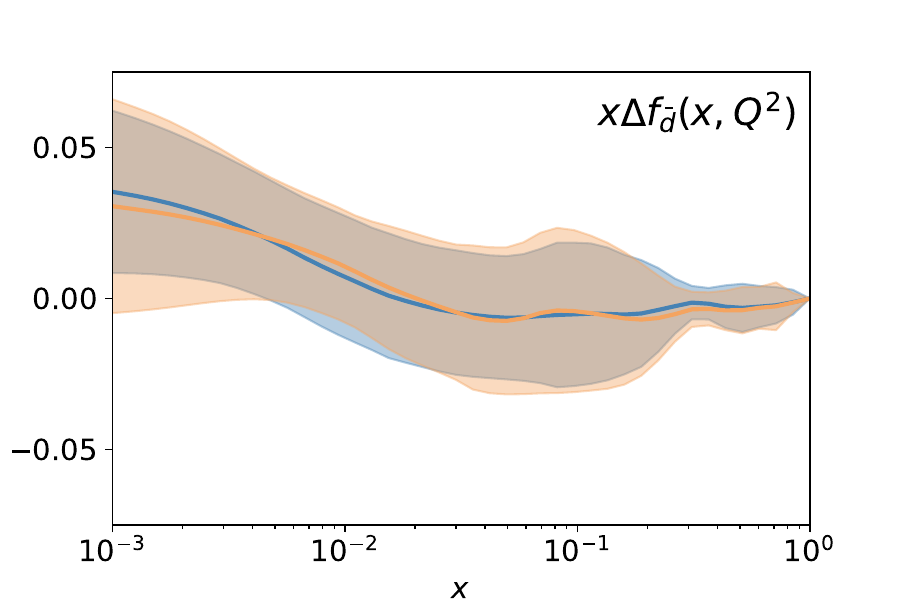}\\
  \includegraphics[width=0.48\textwidth]{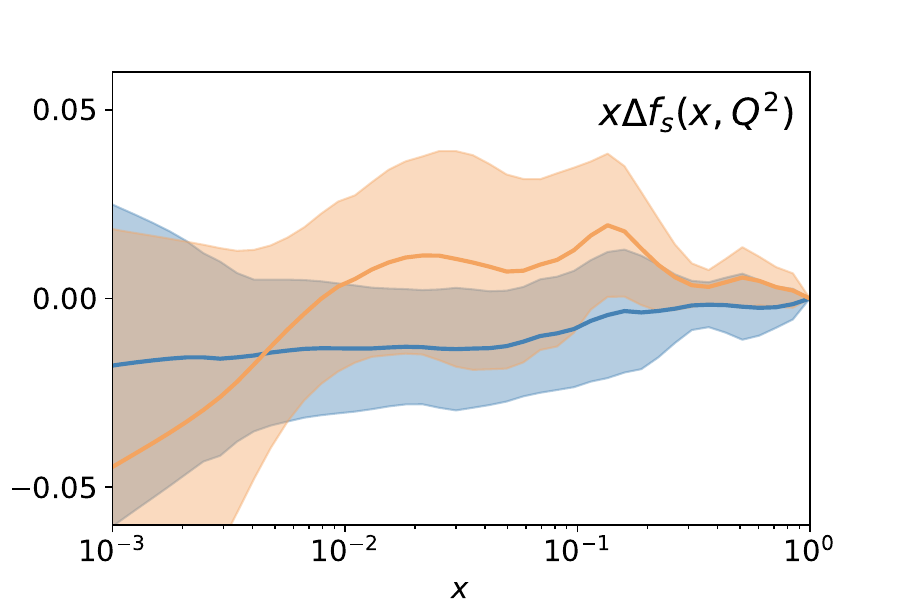}
  \includegraphics[width=0.48\textwidth]{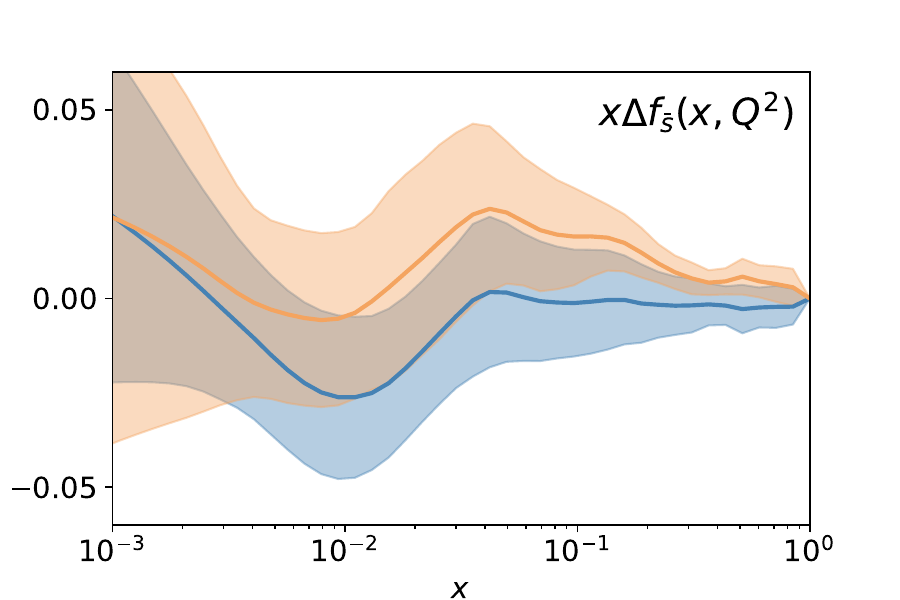}\\
  \includegraphics[width=0.48\textwidth]{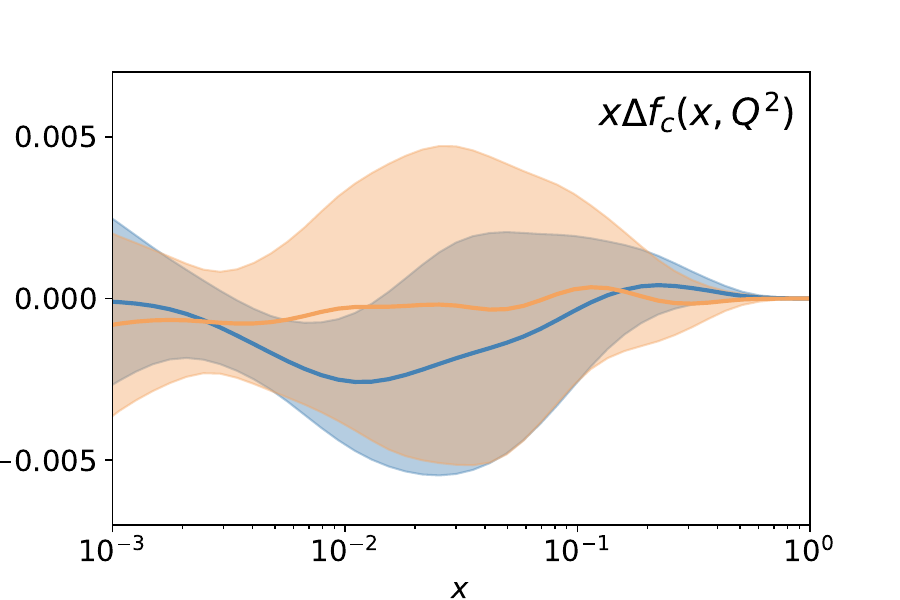}
  \includegraphics[width=0.48\textwidth]{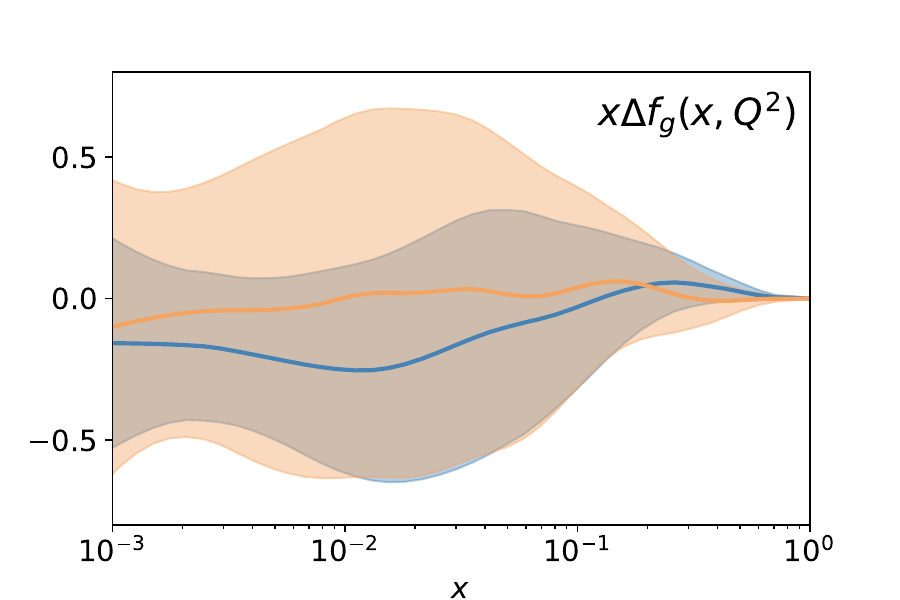}\\
  \vspace{0.5cm}
  \caption{The $\Delta f_u$, $\Delta f_{\bar u}$, $\Delta f_d$, $\Delta f_{\bar d}$,
    $\Delta f_s$, $\Delta f_{\bar s}$, $\Delta f_c$, and $\Delta f_g$ PDFs
    as functions of $x$ at $Q^2=10$~GeV$^2$ from the {\sc MAPPDFpol1.0} NLO
    and NNLO PDF sets. Error bands correspond to one-sigma uncertainties.}
  \label{fig:PDFs_nnlo_vs_nlo}
\end{figure}

In Fig.~\ref{fig:PDFs_nnlo_vs_nlo}, we display the $\Delta f_u$,
$\Delta f_{\bar u}$, $\Delta f_d$, $\Delta f_{\bar d}$, $\Delta f_s$,
$\Delta f_{\bar s}$, $\Delta f_c$, and $\Delta f_g$ PDFs from
the {\sc MAPPDFpol1.0} NNLO and NLO PDF sets as functions of $x$ at
$Q^2=10$~GeV$^2$. Error bands correspond to one-sigma
uncertainties. The impact of NNLO corrections on PDF central
values is generally small in comparison to their uncertainties:
$\Delta f_u$ is enhanced by about half a sigma around $x\sim 0.3$;
$\Delta f_{\bar u}$ is also enhanced by about half a sigma around
$x\sim 0.1$; $\Delta f_s$ and $\Delta f_{\bar s}$ are suppressed by
slightly less than one sigma for $x\gtrsim 0.01$; and $\Delta f_c$ and
$\Delta f_g$ are also suppressed by half a sigma around $x\sim 0.1$.
Uncertainties remain almost unaffected by the inclusion of NNLO
corrections, except in the case of $\Delta f_g$ (and similarly of
$\Delta f_c$), for which they reduce slightly.

In general, the fitted data set is not able to constrain all the
parametrised PDFs to the same level of precision. While DIS data
constrains $\Delta f_u$ and $\Delta f_d$ rather well, SIDIS data
provides limited input on quark flavour separation:
$\Delta f_{\bar u}$ and $\Delta f_{\bar d}$ remain compatible with
zero within uncertainties at both NLO and NNLO; $\Delta f_s$ and
$\Delta f_{\bar s}$ are also compatible with zero within uncertainties
and very similar to each other at both NLO and NNLO. In this last
respect, we conclude that the currently available SIDIS data, even
when analysed including NNLO corrections, is unable to pin down a
possible polarised strange asymmetry. Finally, DIS and SIDIS data
leave $\Delta f_g$ and $\Delta f_c$, the latter generated by
gluon splitting in the perturbative evolution, almost
unconstrained. Indeed, $\Delta f_g$ only enters the DIS and SIDIS
cross sections beyond leading order, therefore it is suppressed by a
power of $\alpha_s$ with respect to the quark distributions. Moreover, scale
violations due to PDF evolution, to which $\Delta f_g$ is
sensitive, are also small because of the limited coverage in $Q^2$ of
the data.

\subsection{Impact of data}
\label{subsec:data}

In order to investigate the relationship between data sets and fit
quality more closely, we assessed the relative impact of DIS and SIDIS
data on PDFs by performing three pairs (NLO and NNLO) of variant fits
to reduced data sets: a first pair from which we removed the COMPASS
SIDIS data; a second pair from which we removed the HERMES SIDIS data;
and a third pair from which we removed the SIDIS data altogether. In
this last case, because the inclusive DIS data alone is not sensitive to
all of the PDF combinations in Eq.~\eqref{eq:basis}, we used the
restricted parametrisation basis
$\left\{\Delta f_{u}^+, \Delta f_{d}^+, \Delta f_{s}^+, \Delta f_g
\right\}$, where $\Delta f_{q}^+=\Delta f_q+\Delta f_{\bar q}$, with
$q=u,d,s$.

\begin{figure}[!t]
  \centering
  \includegraphics[width=\textwidth]{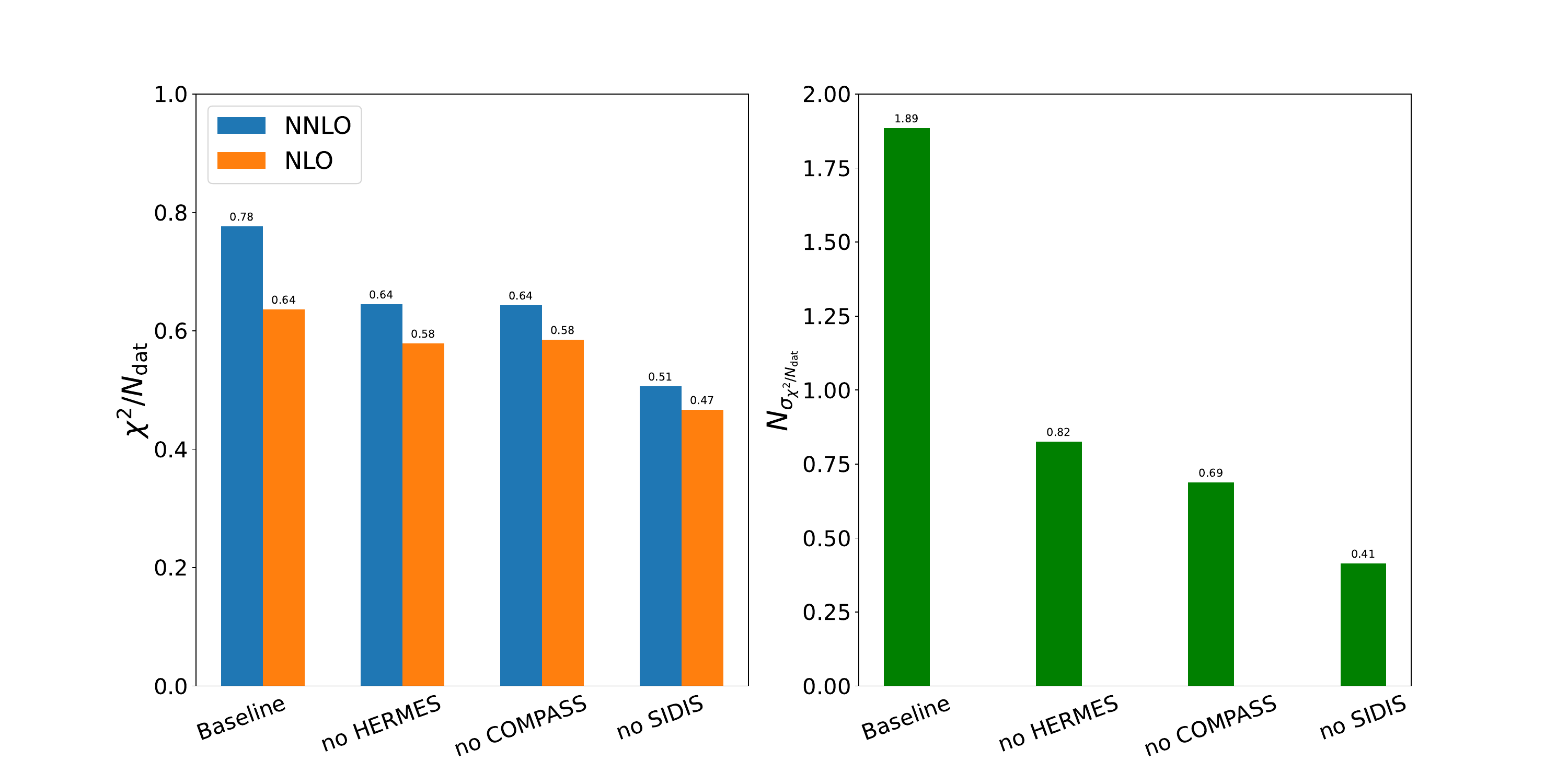} 
  \caption{Left: The values of the global $\chi^2$ per data point for
    the NLO and NNLO baseline fits compared to the corresponding
    values obtained excluding the HERMES SIDIS data, the COMPASS SIDIS
    data, and all SIDIS data. Right: The corresponding fraction of
    standard deviations in units of the $\chi^2$ per data point of the
    difference between the NLO and NNLO $\chi^2$, $N_{\sigma_{\chi^2}/N_{\rm dat}}$.}
  \label{fig:chi2_discr}
\end{figure}

In Fig.~\ref{fig:chi2_discr}, we display, for all of these fits and
for the baseline fits, the values of the global $\chi^2$ per data
point (left), and the fraction of standard deviations in units of
$\chi^2$ per data point corresponding to the difference between NLO
and NNLO, defined as
$N_{\sigma_{\chi^2}/N_{\rm dat}}=(\chi^2_{\rm NNLO}-\chi^2_{\rm NLO})/\sqrt{2N_{\rm dat}}$
(right).  In all cases, the global $\chi^2$ remains larger at NNLO than at NLO.
However, $N_{\sigma_{\chi^2}/N_{\rm dat}}$ reduces from about two to less than
one when either the HERMES or the COMPASS SIDIS measurements are
removed from the fit, and to one half when all SIDIS data is removed
altogether from the fit. Therefore, in these three last cases the
significance of the increase in $\chi^2$ observed when including NNLO
corrections is compatible with a one-sigma statistical fluctuation of
the $\chi^2$.  This analysis also suggests that there are two effects
responsible for the increase of the $\chi^2$ per data point upon inclusion of
NNLO corrections: the fact that the SIDIS data sets are not
described as well as the DIS ones; and the fact that HERMES and COMPASS
SIDIS data sets, while being equally well described separately, are no
longer so when included together in the fit. The latter fact, which may
point towards an inconsistency between the two experiments, seems to
lead to the largest increase of $\chi^2$.

\begin{figure}[!t]
  \centering
  \includegraphics[width=0.48\textwidth]{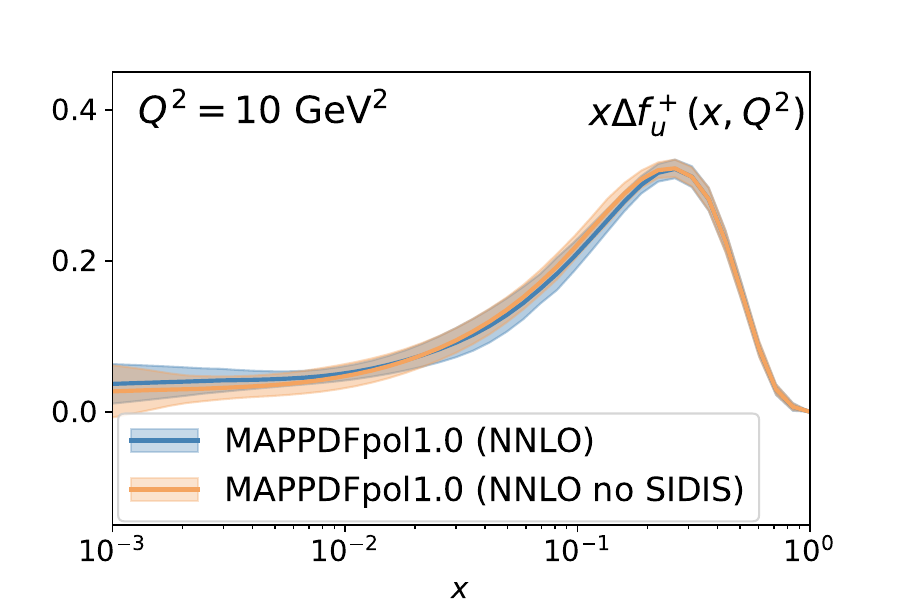}
  \includegraphics[width=0.48\textwidth]{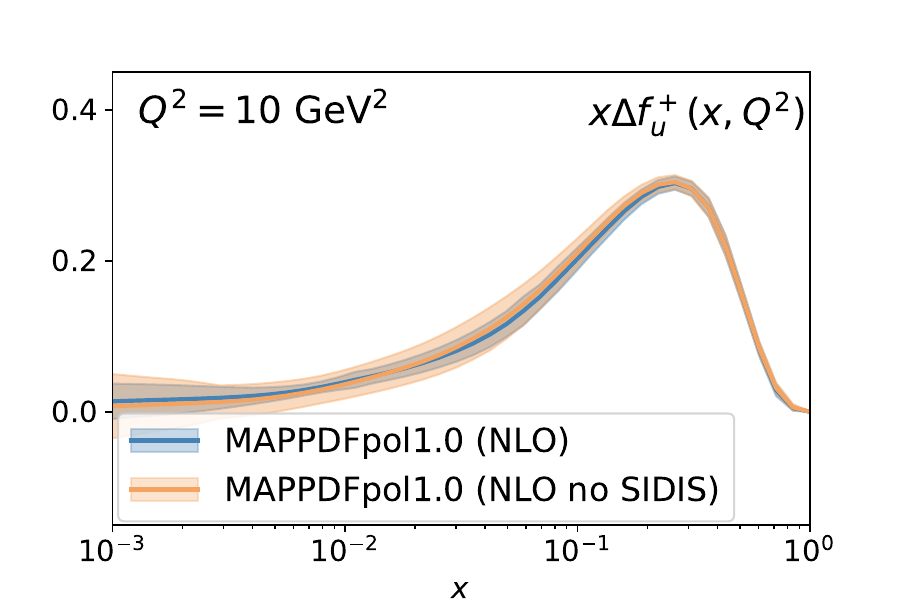}\\
  \includegraphics[width=0.48\textwidth]{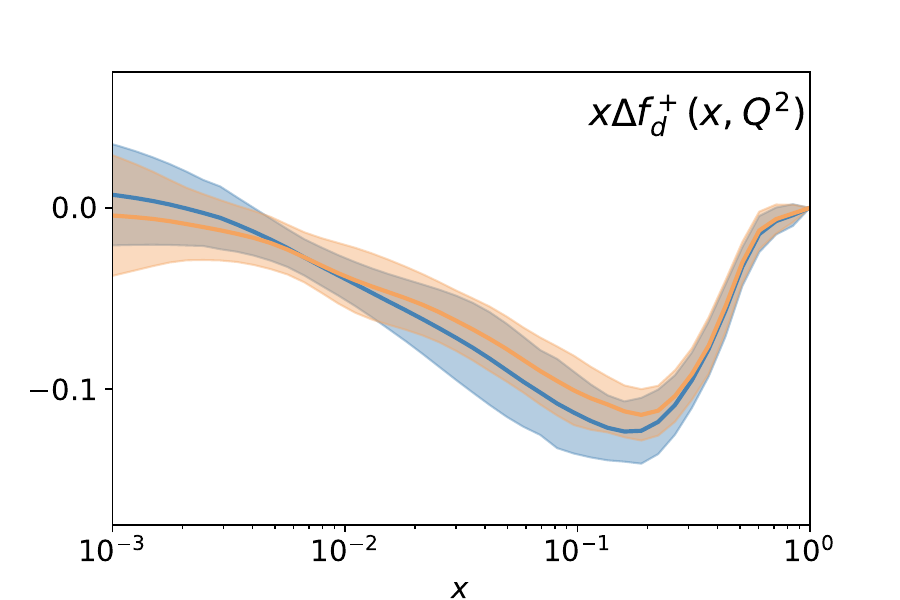}
  \includegraphics[width=0.48\textwidth]{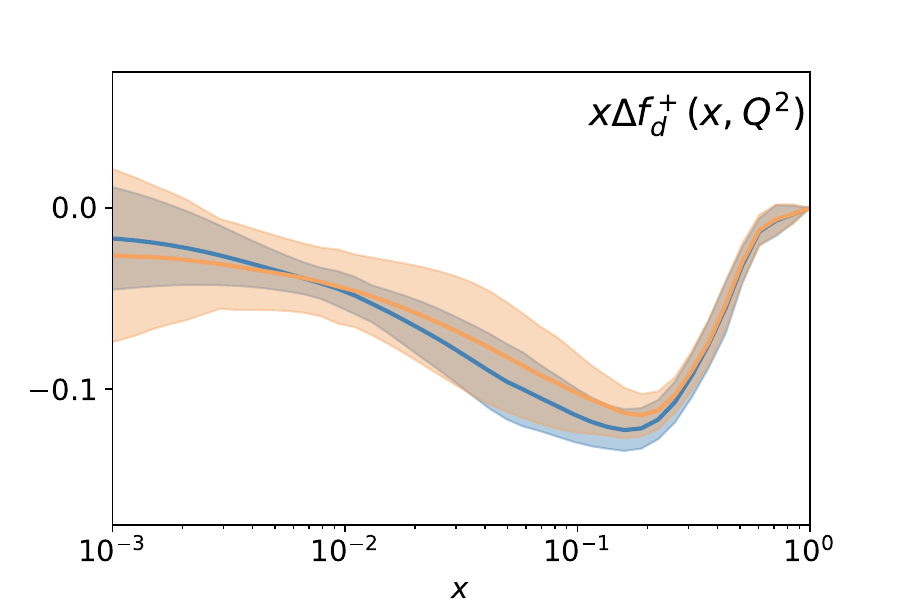}\\
  \includegraphics[width=0.48\textwidth]{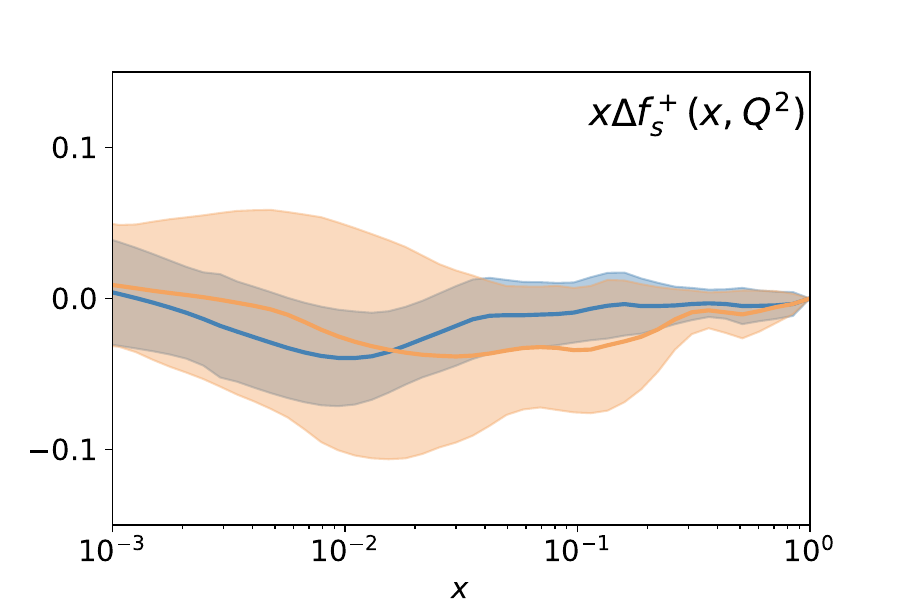}
  \includegraphics[width=0.48\textwidth]{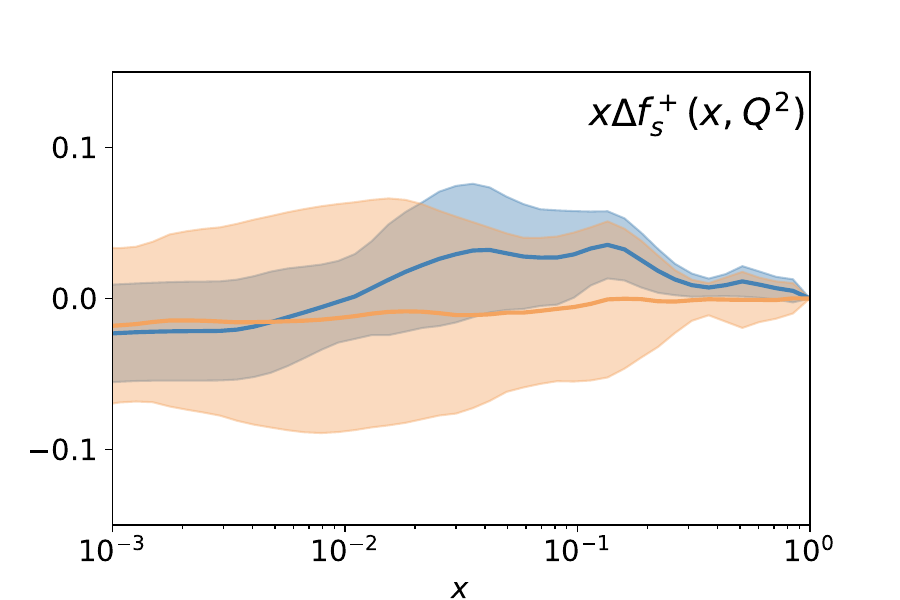}\\
  \includegraphics[width=0.48\textwidth]{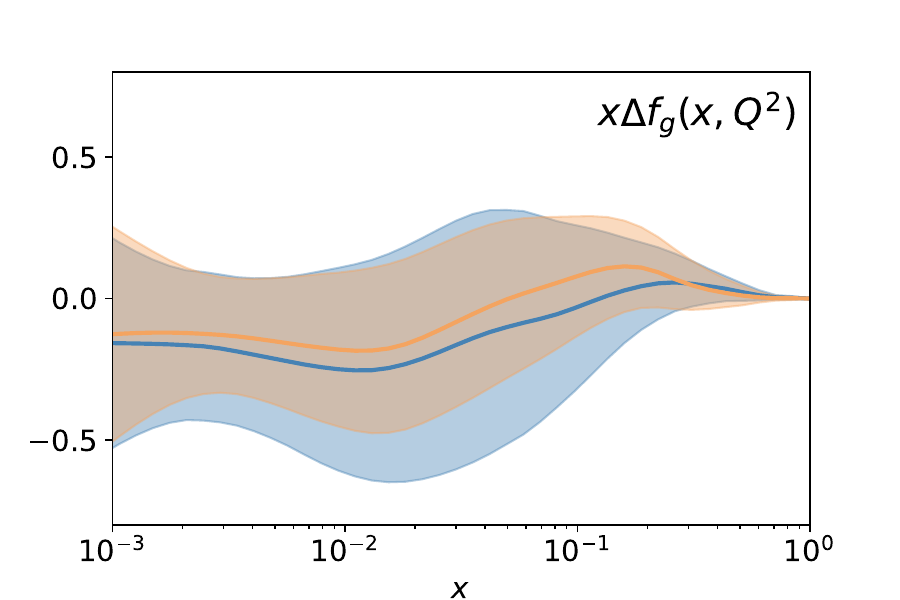}
  \includegraphics[width=0.48\textwidth]{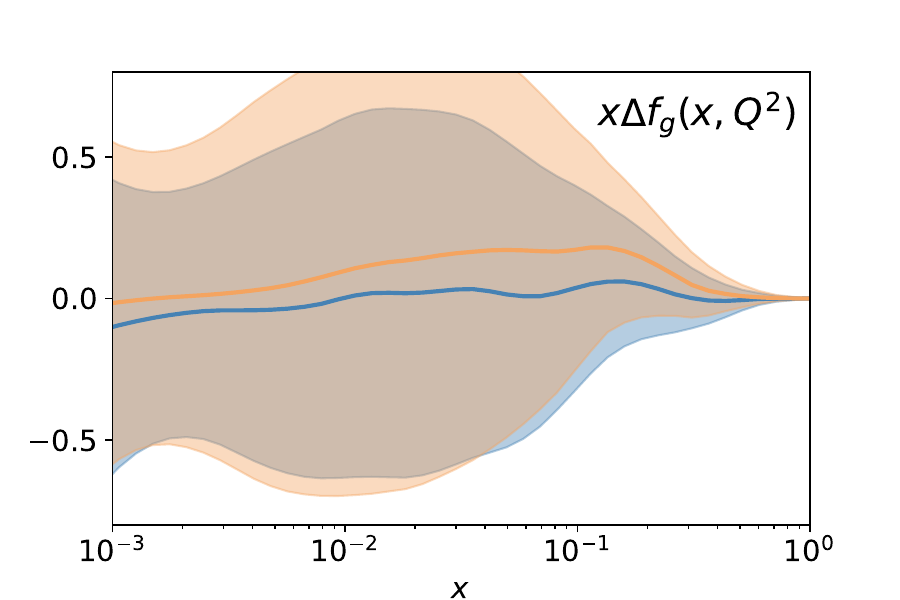}\\
  \vspace{0.5cm}
  \caption{The $\Delta f_{u}^+$, $\Delta f_{d}^+$, $\Delta f_{s}^+$,
    and $\Delta f_g$ PDF combinations as functions of $x$ at
    $Q^2=10$~GeV$^2$ from the {\sc MAPPDFpol1.0} NNLO (left) and NLO
    (right) PDF sets compared to the PDFs from the corresponding sets
    without SIDIS data. Error bands correspond to one-sigma
    uncertainties.}
  \label{fig:PDFs_data}
\end{figure}

In Fig.~\ref{fig:PDFs_data}, we compare the
$\Delta f_{u}^+$, $\Delta f_{d}^+$, $\Delta f_{s}^+$, and $\Delta f_g$ PDF
combinations obtained in the baseline fits and in the fits with no SIDIS data
included. The comparison is reported at $Q^2=10$~GeV$^2$ for both NNLO and NLO.
Error bands correspond to one-sigma uncertainties. We observe a non-trivial
interplay between SIDIS data and NNLO corrections.

In the case of $\Delta f_{s}^+$, SIDIS data has a consistent impact at
NLO and NNLO, which results in a reduction of PDF uncertainties by up
to 50\% for $0.01\lesssim x\lesssim 0.1$. Interestingly,
$\Delta f_{s}^+$ turns out to be compatible in the global
determination and in the determination without SIDIS data. This is in
contrast with previous analyses (see {\it e.g.}~\cite{Leader:2014uua}
and references therein) in which a tension between DIS and SIDIS data
was claimed, with the former leading to a markedly negative
$\Delta f_{s}^+$ and the latter to a sign-changing $\Delta f_{s}^+$.
We instead find results compatible with zero within large
uncertainties in both cases, which are even more so upon inclusion of
NNLO corrections. We believe that this is a consequence of the
combination of our parametrisation, which is more flexible than that
used in previous analyses, and of the usage of the {\sc MAPFF1.0} kaon
FFs~\cite{AbdulKhalek:2022laj}, which were determined with the same
methodology used to determine the current PDFs.

In the case of $\Delta f_{u}^+$ and of $\Delta f_{d}^+$, SIDIS data
has a different impact at NLO and NNLO. At NLO, we observe a reduction
of the uncertainties for all values of $x$; at NNLO, instead, we
observe an increase of the uncertainties for all values of $x$. We
therefore conclude that the inclusion of NNLO corrections somewhat
amplifies an underlying tension in the SIDIS data sets. Given that
this behaviour is observed for $\Delta f_{u}^+$ and $\Delta f_{d}^+$,
but not for $\Delta f_{s}^+$, we conclude that the pion data and
FFs need further investigation, which we leave for a future
study. Finally, as expected, $\Delta f_g$ is left unaltered by SIDIS
data at both NLO and NNLO.

\begin{figure}[!t]
  \centering
  \includegraphics[width=0.49\textwidth]{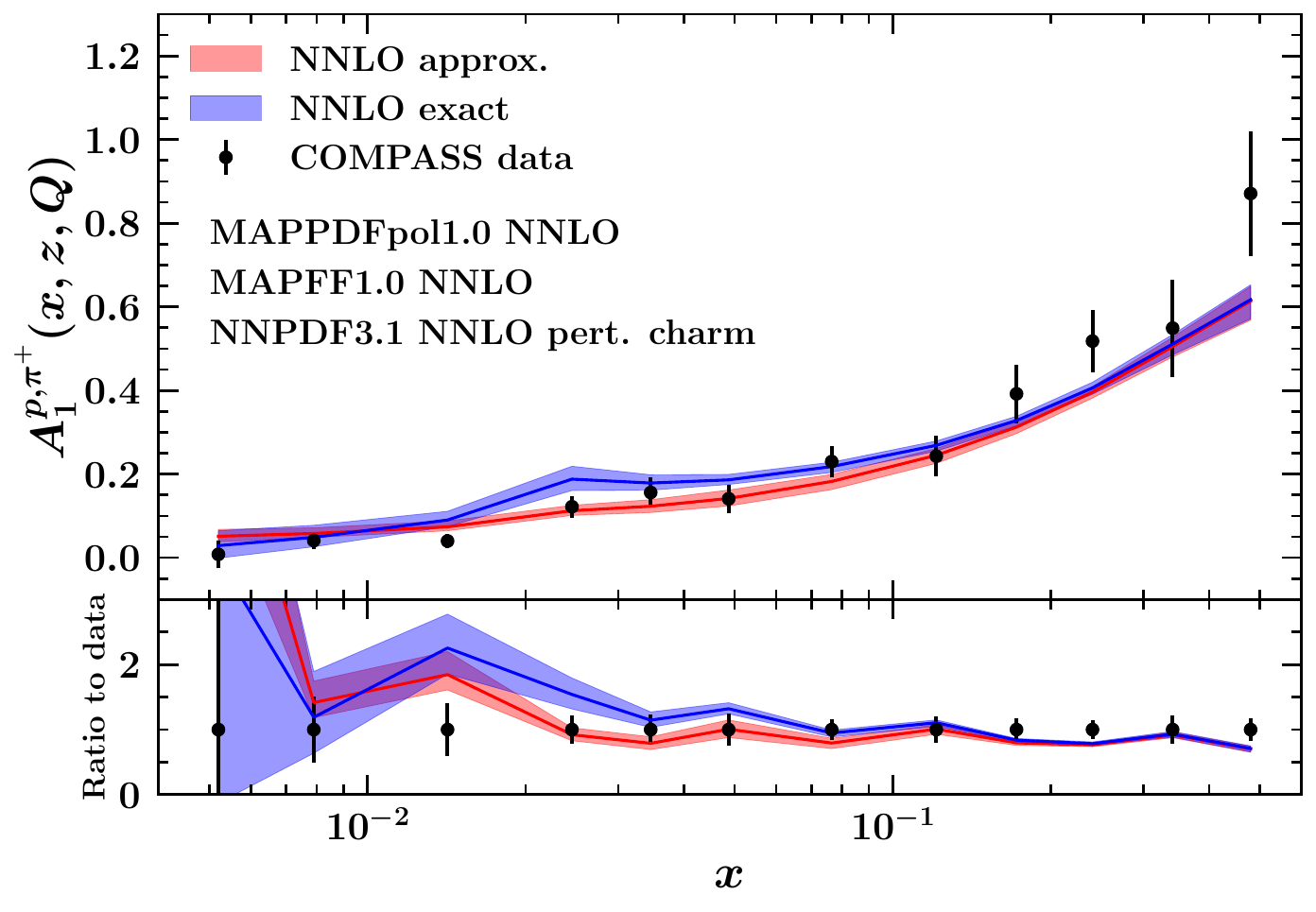}
  \includegraphics[width=0.49\textwidth]{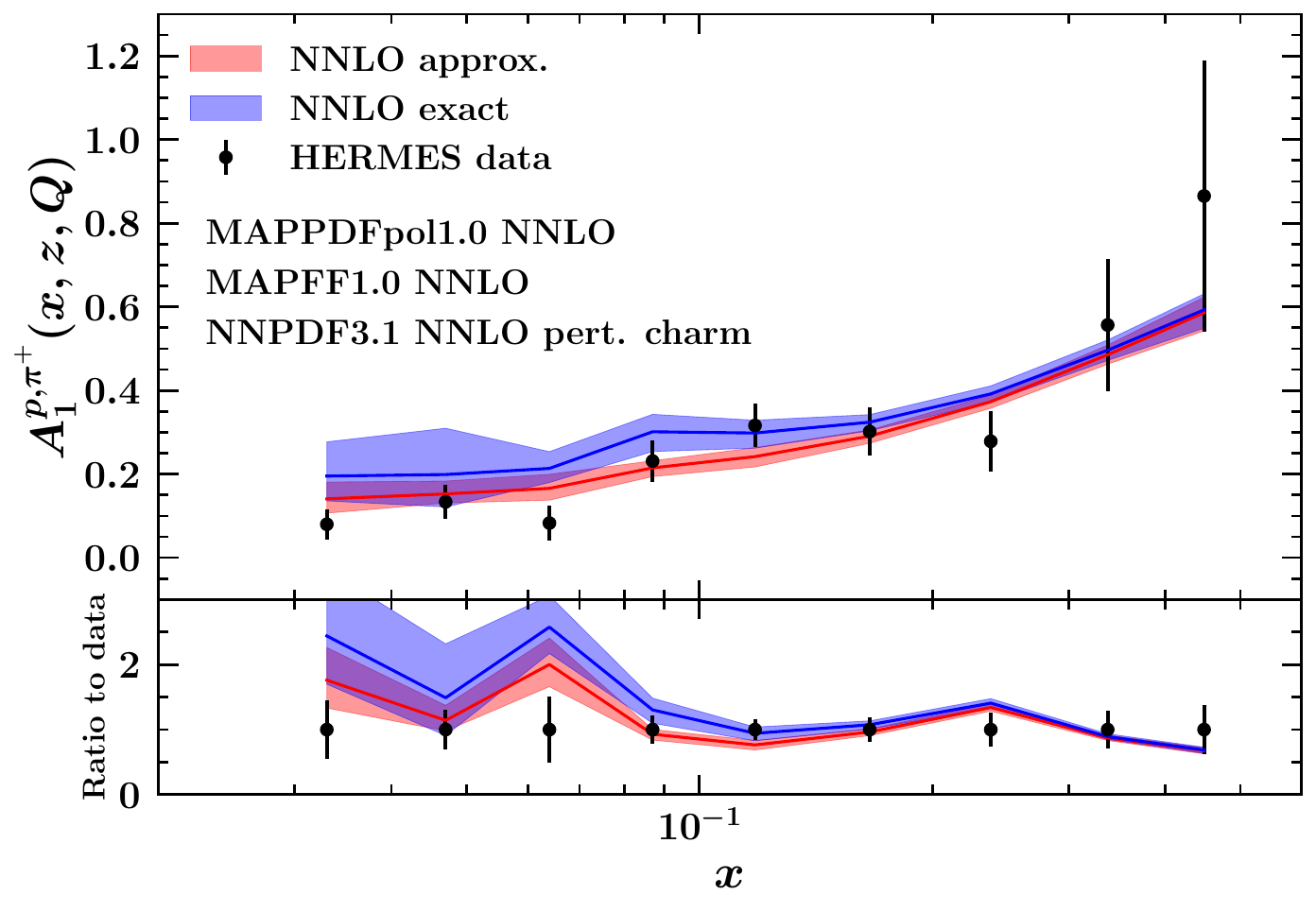}\\
  \includegraphics[width=0.49\textwidth]{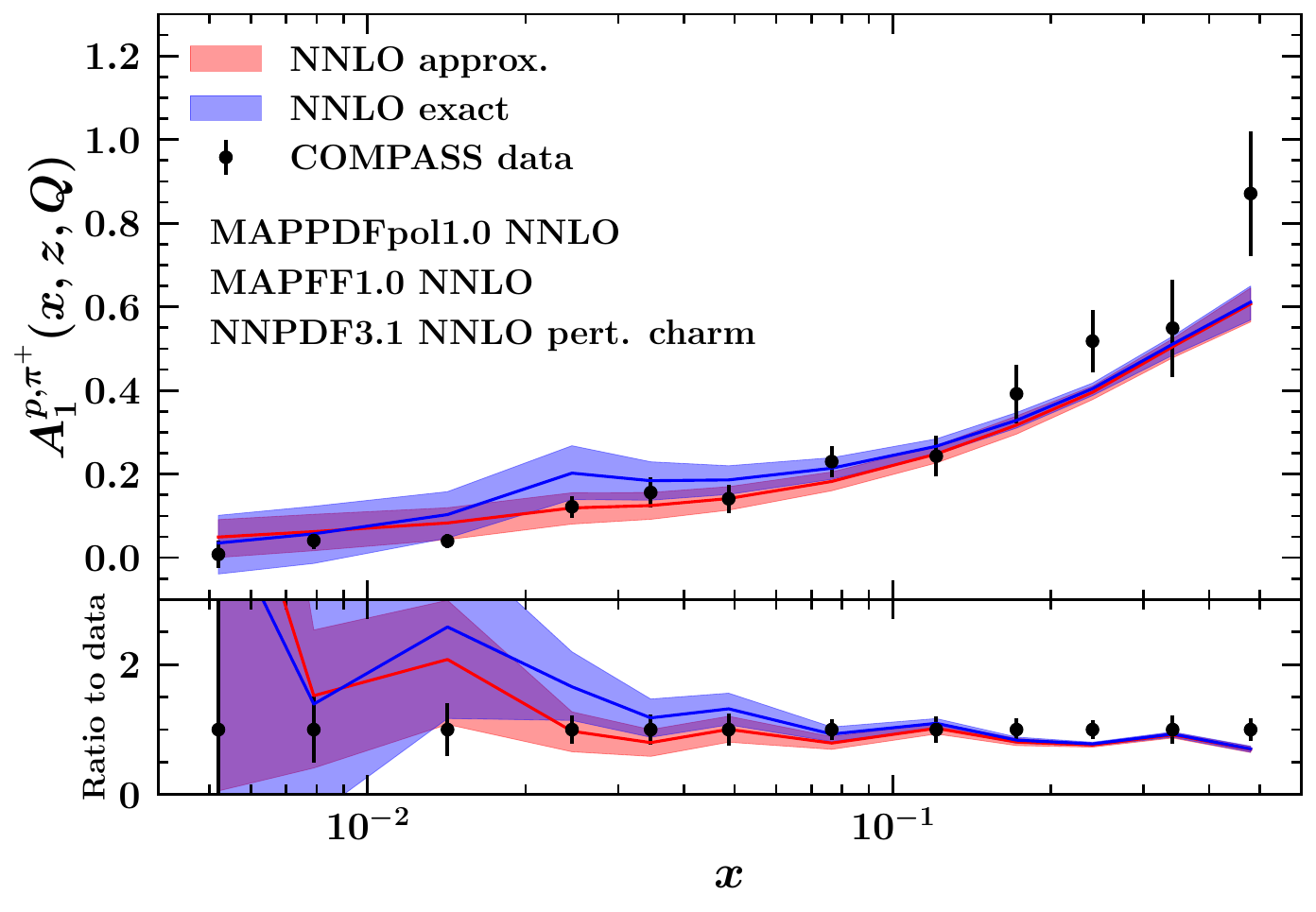}
  \includegraphics[width=0.49\textwidth]{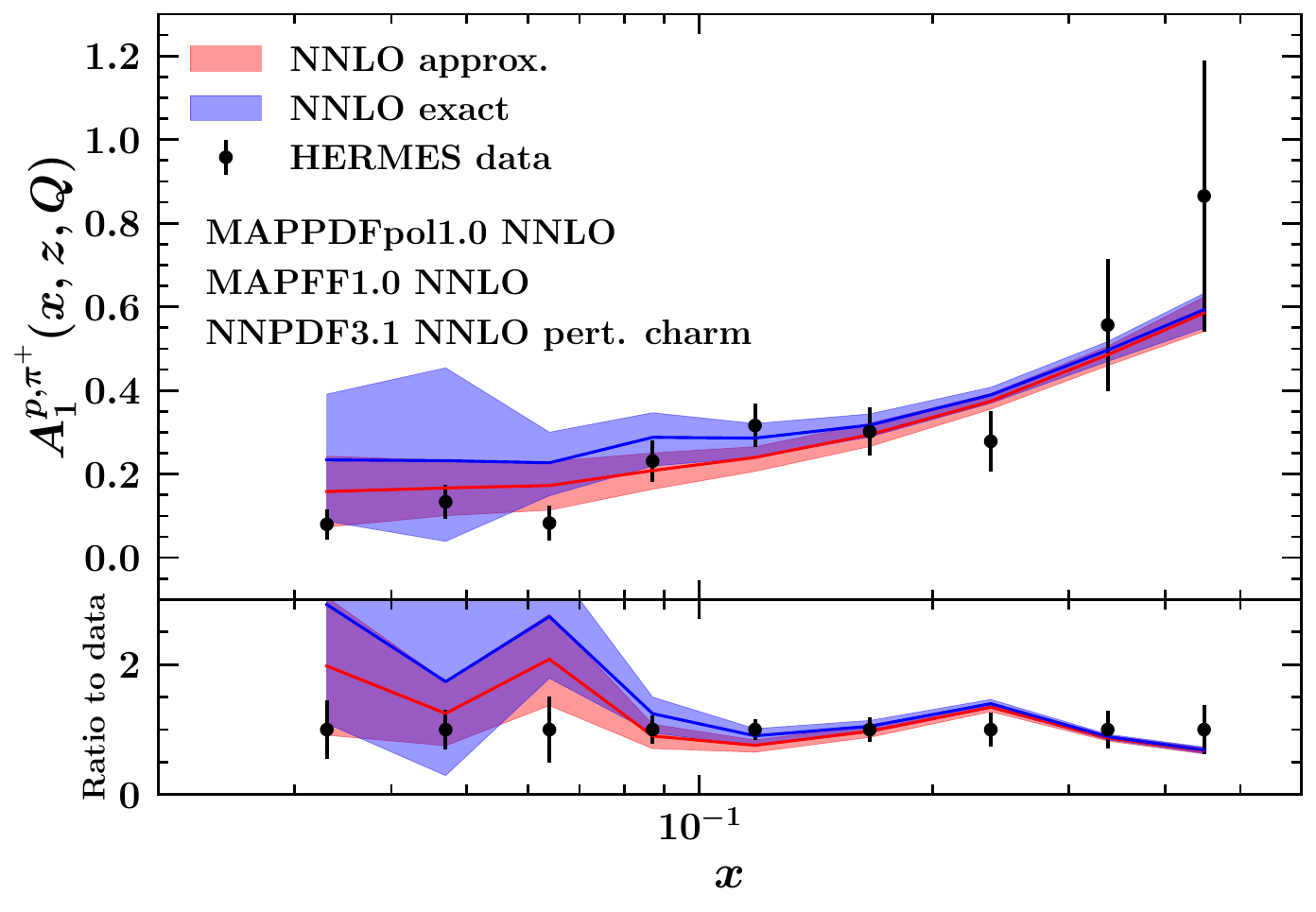}\\  
  \vspace{0.5cm}
  \caption{Comparison between experimental data and theoretical predicitons
    of representative measurements of the SIDIS single-spin asymmetry
    $A_1^{p,\pi^+}=g_1^{p,\pi^+}/F_1^{p,\pi^+}$ performed by COMPASS (left) and
    HERMES (right) experiments. Theoretical predictions are computed with the
    {\sc NNPDF3.1} NNLO unpolarised PDFs, the {\sc MAPFF1.0} NNLO FFs using
    and the default {\sc MAPPDFpol1.0} NNLO polarised PDFs (upper plots) or
    the {\sc MAPPDFpol1.0} NNLO polarised PDFs wiht  restricted cut on SIDIS
    data. In each plot, the two curves correspond to results obtained using
    either the approximate~\cite{Abele:2021nyo} or the
    exact~\cite{Bonino:2024wgg,Goyal:2024tmo} computations of NNLO corrections
    to SIDIS matrix elements.}
  \label{fig:datatheory}
\end{figure}

Finally, in order to investigate how much of the observations made above come
from the fact that we are analysing SIDIS data with approximate NNLO
corrections, we compare some representative COMPASS and HERMES SIDIS
measurements with two sets of theoretical predictions. One set is obtained with
the approximate computation of NNLO corrections to SIDIS matrix
elements~\cite{Abele:2021nyo}, the other set with the exact
computation~\cite{Bonino:2024wgg,Goyal:2024tmo}, which we have carefully
benchmarked against Ref.~\cite{Bonino:2024wgg}. Both sets are obtained
with fixed input PDFs and FFs: the default NNLO polarised PDFs determined in
Sect.~\ref{subsec:nnlo}, and the {\sc NNPDF3.1} NNLO unpolarised PDFs
and {\sc MAPFF1.0} NNLO FFs used in the corresponding fit. The comparison
between data and theory is displayed in the upper panels of
Fig.~\ref{fig:datatheory}.

As we can see form Fig.~\ref{fig:datatheory}, the two theoretical computations
are in good agreement, within uncertainties, except for COMPASS around values
of $x\sim 0.03$. The peculiar bump of the asymmetry obtained with the
exact computation can potentially be reabsorbed in the PDF once this is
refitted using the same predictions. This exercise, however, goes beyond the
scope of this work: for consistency we would have to refit the input FFs,
using the exact NNLO computation also for unpolarised SIDIS.

Nevertheless, we can test the robustness of our polarised PDFs by repeating the
baseline NLO and NNLO determinations without the SIDIS data in the kinematic
region where the approximate and exact predictions differ by an amount larger
than the experimental uncertainty. Specifically, we repeat the baseline NLO and
NNLO fits applying a further cut to COMPASS and HERMES SIDIS measurements, in
which we retain only the data points that satify $x\geq x_{\rm cut}$, with
$x_{\rm cut}=0.1$. In this case, the fit quality of the new NLO and NNLO fits is
almost the same, and very close to that of the baseline NLO fit reported in
Table~\ref{tab:chi2}. The shift of the central value of the NNLO polarised PDF
set obtained with the more restrictive cut with respect to the central value of
the baseline NNLO polarised PDF set is mild, and well encompassed by the
uncertainties of the latter set. Theoretical predictions, obtained with either
the approximate or exact computations and with the new NNLO fit with the more
restrictive cut are displayed in the lower panels of Fig.~\ref{fig:datatheory}.
As we can see, the peculiar bump  of the asymmetry obtained with the exact
computation remains, though this is now accompanied by a more conservative
uncertainty that makes it compatible with the asymmetry obtained with the
approximate computation. All of these results are consistent with those
reported in Ref.~\cite{Borsa:2024mss}.

In light of these considerations, we conclude that refitting polarised PDFs
using the exact computation for polarised SIDIS may indeed reduce the global
$\chi^2$ of the fit, though it will modify polarised PDFs only mildly.
An explicit refit is left to future work, in which we will also consistently
refit, with the exact computation, the FFs used as input to the determination
of the polarised PDFs. In this case, because new (gluon-initiated) channels
open up at NNLO, we expect some significant changes in the gluon FF. These
changes would however only mildly affect a fit of polarised PDFs. The reason
being that COMPASS and HERMES experiments measure asymmetries,
that is the ratio between the polarised and unpolarised SIDIS structure
functions, in which the FF dependence largely cancels out.

\subsection{Impact of theoretical constraints}
\label{subsec:theory}

\begin{figure}[!t]
  \centering
  \includegraphics[width=0.48\textwidth]{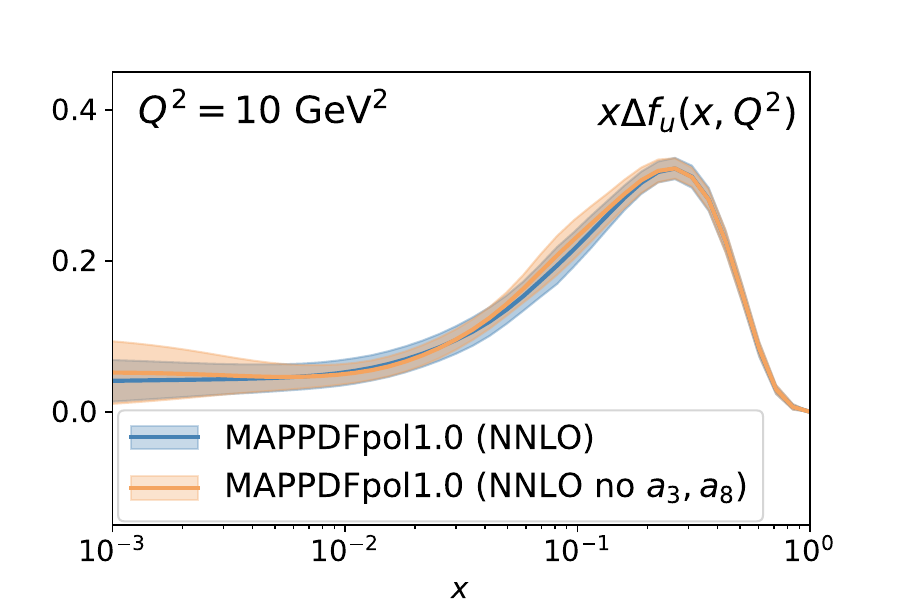}
  \includegraphics[width=0.48\textwidth]{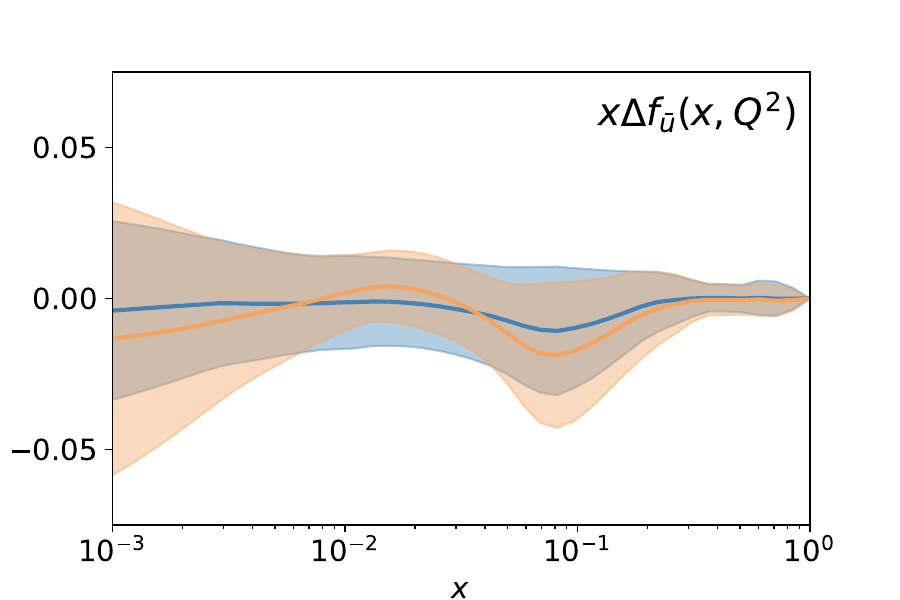}\\
  \includegraphics[width=0.48\textwidth]{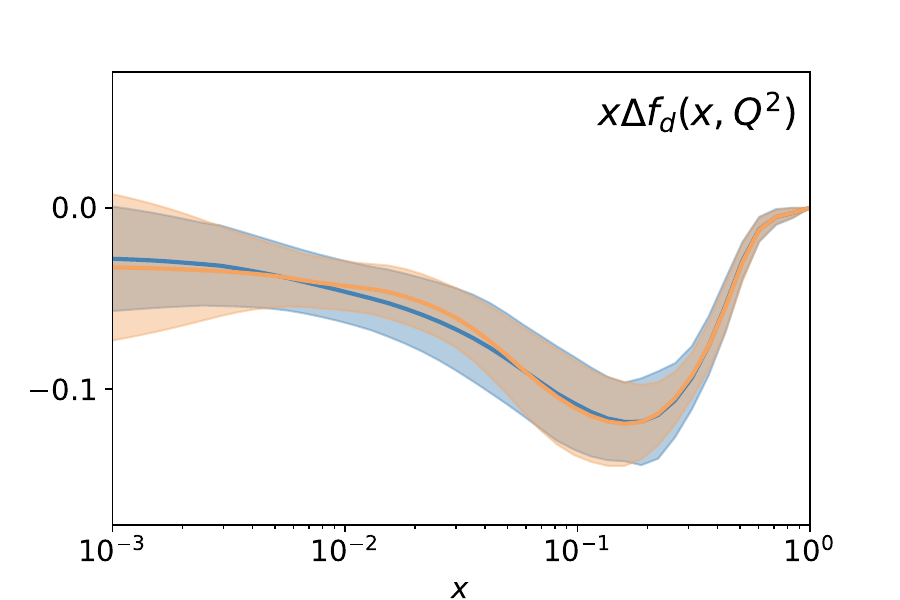}
  \includegraphics[width=0.48\textwidth]{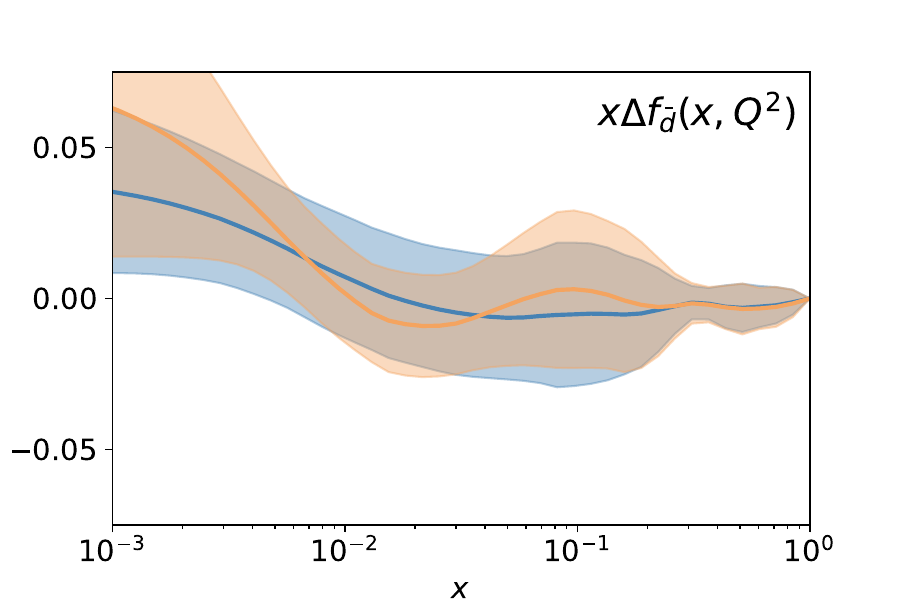}\\
  \includegraphics[width=0.48\textwidth]{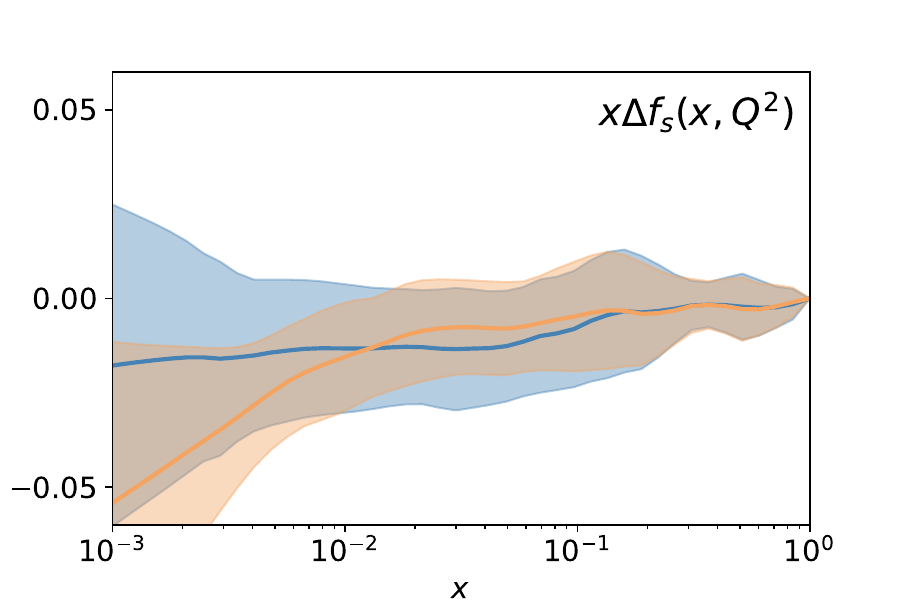}
  \includegraphics[width=0.48\textwidth]{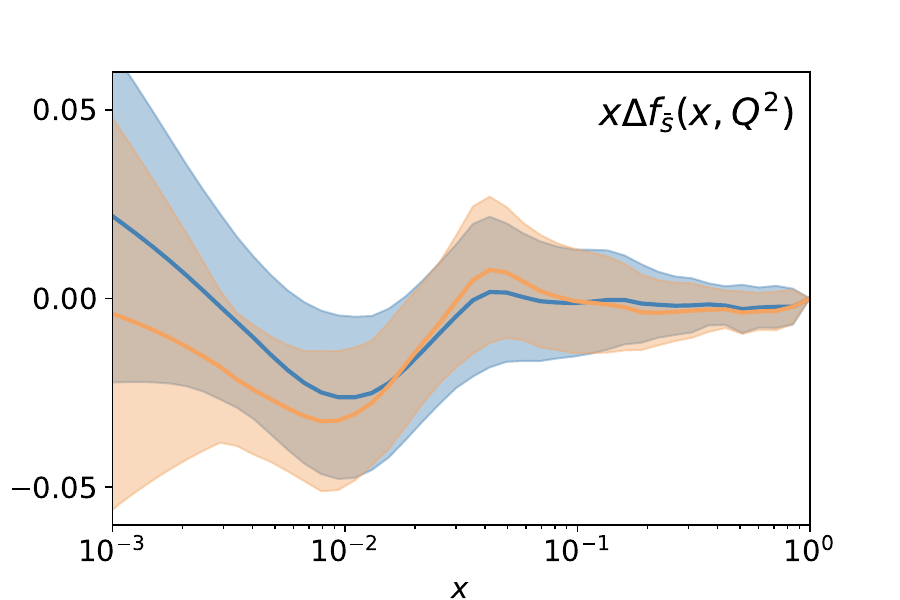}\\
  \vspace{0.5cm}
  \caption{The $\Delta f_u$, $\Delta f_{\bar u}$, $\Delta f_d$,
    $\Delta f_{\bar d}$, $\Delta f_s$, and $\Delta f_{\bar s}$ PDFs as
    functions of $x$ at $Q^2=10$~GeV$^2$ from the {\sc MAPPDFpol1.0}
    NNLO PDF sets with and without data for $a_3$ and $a_8$
    included. Error bands correspond to one-sigma uncertainties.}
  \label{fig:PDFs_no_a3_a8}
\end{figure}

\begin{figure}[!t]
  \centering
  \includegraphics[width=0.48\textwidth]{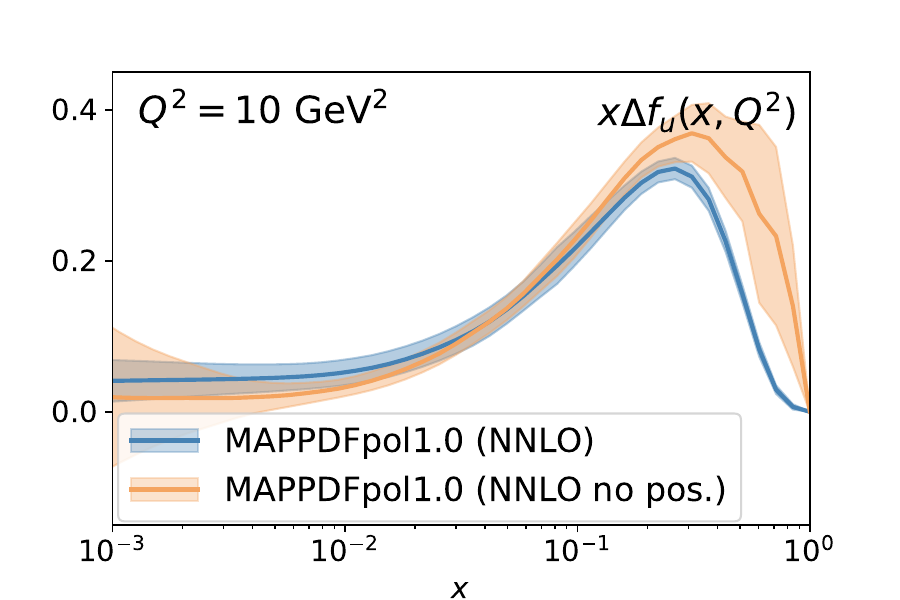}
  \includegraphics[width=0.48\textwidth]{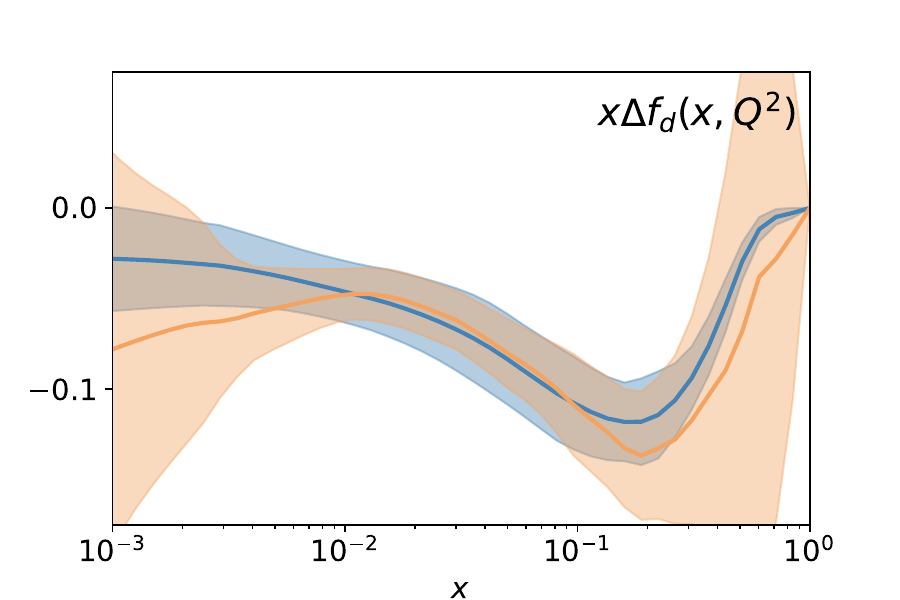}\\
  \includegraphics[width=0.48\textwidth]{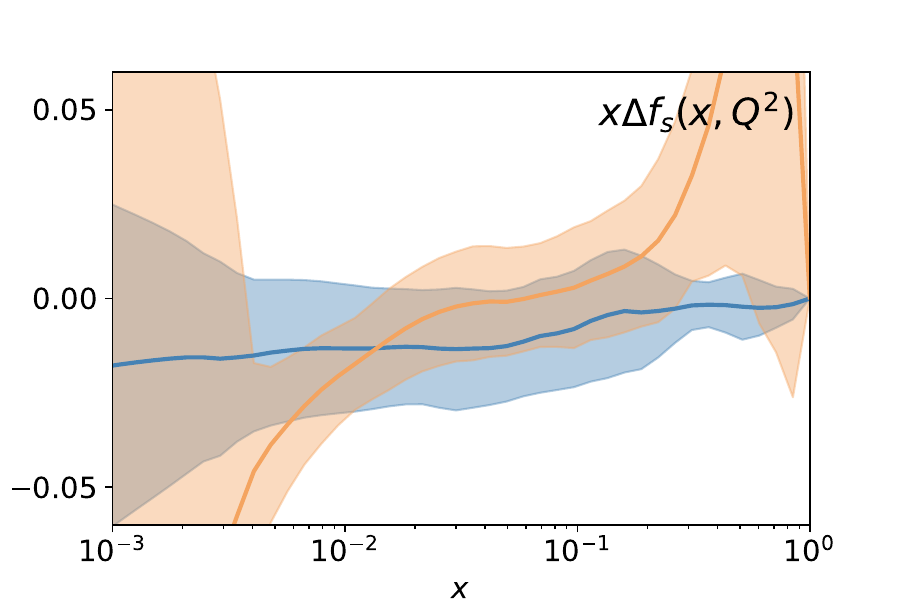}
  \includegraphics[width=0.48\textwidth]{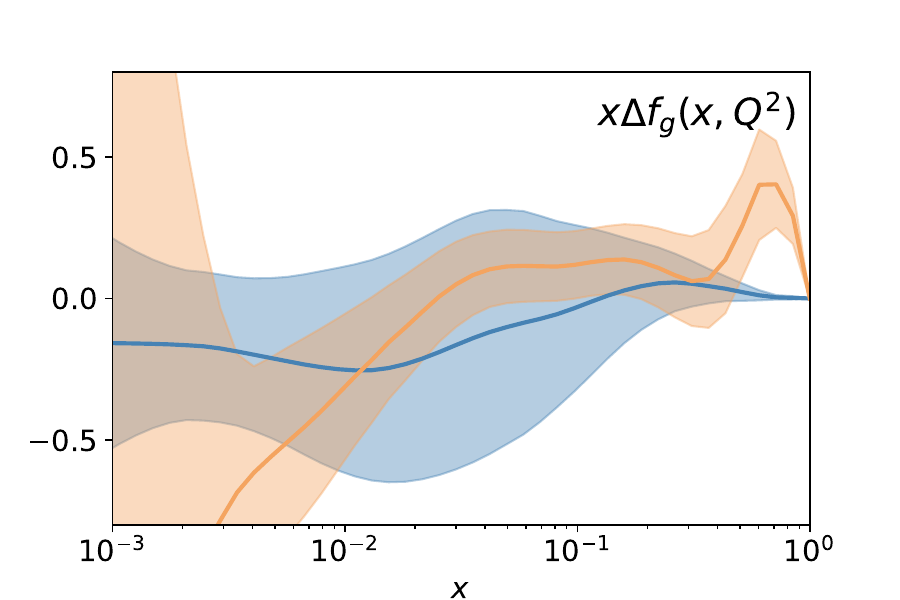}\\
  \vspace{0.5cm}
  \caption{The $\Delta f_u$, $\Delta f_d$, $\Delta f_s$, and $\Delta f_g$,
    PDFs as functions of $x$ at $Q^2=10$~GeV$^2$ from the {\sc MAPPDFpol1.0}
    NNLO PDF sets with and without the positivity constraint imposed.
    Error bands correspond to one-sigma uncertainties.}
  \label{fig:PDFs_nopos}
\end{figure}

We finally investigate the impact of including data for $a_3$ and
$a_8$ and of imposing the positivity constraint.  To this purpose, we
performed three additional fits: two fits, one at NLO and one at NNLO,
from which we removed the data points for $a_3$ and $a_8$; and one NNLO fit
in which we set $K=50$ in Eq.~\eqref{eq:pos_net_alt}. This choice
makes the positivity constraint so loose that it virtually corresponds
to removing it altogether. The fit quality of each of these additional
fits, as quantified by the $\chi^2$ per data point, is reported in
Table~\ref{tab:chi2} for the individual and for the global data
sets. The corresponding NNLO PDFs are displayed in
Figs.~\ref{fig:PDFs_no_a3_a8} and \ref{fig:PDFs_nopos}:
in Fig.~\ref{fig:PDFs_no_a3_a8}, we compare the
$\Delta f_u$, $\Delta f_{\bar u}$, $\Delta f_d$,
$\Delta f_{\bar d}$, $\Delta f_s$, and $\Delta f_{\bar s}$ PDFs from
the {\sc MAPPDFpol1.0} NNLO PDF sets with and without $a_3$ and $a_8$;
in Fig.~\ref{fig:PDFs_nopos}, we compare the $\Delta f_u$, $\Delta f_d$,
$\Delta f_s$, and $\Delta f_g$ PDFs from the {\sc MAPPDFpol1.0} NNLO PDF sets
with and without the positivity constraint imposed. All comparisons are
displayed as functions of $x$ at $Q^2=10$~GeV$^2$. Error bands
correspond to one-sigma uncertainties.

Concerning the impact of the $a_3$ and $a_8$ data points, we see that it is
negligible. The fit quality remains almost unaltered at both NLO and NNLO
and so do PDF central values and uncertainties. We therefore conclude that
the current DIS and SIDIS data does not point towards any significant violation
of the SU(2) and SU(3) flavour symmetries. This finding is consistent
with~\cite{Ethier:2017zbq}, where polarised PDFs and FFs were determined at NLO
from a set of data very similar to ours.

Concerning the impact of the positivity constraint, we see that it is
extremely relevant. Whereas the quality of the NNLO fit improves when
removing it (incidentally, to a level close to that of the NLO fit
with the positivity constraint), PDFs display peculiar shapes with
large uncertainties in the large-$x$ region. This is the region where
the data is scarcer and where, therefore, the positivity constraint
becomes dominant.

\section{Summary and outlook}
\label{sec:conclusions}

In this paper, we have presented {\sc MAPPDFpol1.0}, a new determination of the
polarised PDFs of the proton from a global QCD analysis of the available
DIS and SIDIS data. This determination includes NNLO QCD
corrections to the DGLAP evolution equations and to the matrix elements of the
two processes analysed. In the case of SIDIS, NNLO QCD corrections to the
matrix elements are implemented using the approximate computation of
Ref.~\cite{Abele:2021nyo}. This determination has been carried out by extending
the framework that we previously developed to determine the FFs of
pions and kaons~\cite{Khalek:2021gxf,AbdulKhalek:2022laj}. It combines a
neural-network parametrisation of PDFs with a Monte Carlo representation of
their uncertainties, aiming at reducing parametrisation bias as much as
possible and at obtaining statistically sound uncertainties.

Our main findings can be summarised as follows.
\begin{itemize}
  
\item There is a subtle interplay between NNLO corrections and experimental
  data. Whereas all the data sets analysed are generally well described,
  the inclusion of NNLO corrections leads to a two sigma increase of the
  $\chi^2$ per data point. This increase is observed also in
  fits without SIDIS data and in fits without either HERMES or COMPASS SIDIS
  data (at a level of about half a sigma), although the largest deterioration
  in the fit quality is seen when HERMES and COMPASS SIDIS measurements are
  simultaneously included in the fit. This behaviour, already observed when
  determining FFs from unpolarised SIDIS data~\cite{AbdulKhalek:2022laj},
  calls for further investigations both on the experimental side and on the
  accuracy of the approximate computation of NNLO corrections to the SIDIS
  cross sections. In this last respect, it was recently shown that the
  approximate computation of Ref.~\cite{Abele:2021nyo} is well reproduced by
  the exact computation in the unpolarised~\cite{Goyal:2023xfi, Bonino:2024qbh}
  and polarised~\cite{Bonino:2024wgg,Goyal:2024tmo} cases at sufficiently
  large values of $x$. We have verified that the increase of the $\chi^2$
  of SIDIS data can be indeed remedied if one repeats the NLO and NNLO fits
  with a cut $x\geq x_{\rm cut}$, with $x_{\rm cut}=0.1$. The effect on PDFs remains
  however moderate, very well within their uncertainties. This finding
  confirms the robustness of our determination, and is consistent with what was
  reported in Ref.~\cite{Borsa:2024mss}. In a similar spirit, we have also
  compared representative COMPASS and HERMES measurements to theoretical
  predictions determined with fixed input polarised/unpolarised PDFs and FFs,
  using either the approximate or the exact computation. We have found little
  difference between the two. Whether the replacement of the approximate
  computation with the exact computation reduces the $\chi^2$, when no cut on
  $x$ is applied to the data, can be checked only by refitting the PDFs.
  We leave this exercise to future work, in which we will also consistently
  redetermine, using the exact computation, the FFs used as input to the fit of
  the polarised PDFs. We expect the new (gluon-initated) channels accounted for
  by the exact NNLO computation to modify the FFs. However this will possibly
  be inconsequential for polarised PDFs, that are determined from spin
  asymmetries, in which the contribution of FFs cancels out in the ratio.
  The impact of NNLO corrections on polarised PDFs is otherwise
  moderate, as they leave PDF central values almost unchanged, and lead to a
  small reduction of uncertainties for the polarised strangeness and gluon PDFs.

\item The analysed data is not able to constrain all parametrised PDFs
  to the same accuracy. Whereas the up- and down-quark polarised PDFs
  are generally well constrained and the SIDIS data moderately help
  pin down uncertainties, the sea-quark and gluon polarised PDFs
  remain largely unconstrained and essentially compatible with zero
  within their large uncertainties. The combination of SIDIS data for
  kaon production and NNLO corrections is not able to tell whether there is an
  asymmetry between strange quark and strange antiquark polarised PDFs,
  which are parametrised independently in our analysis for the first time.
  However, SIDIS data helps reduce uncertainties on these PDFs significantly.
  Noteworthy is the fact that the total polarised strange PDF
  turns out to be similar in the global determination and in the
  determination without SIDIS data, and in turn compatible with zero
  within uncertainties. This is in contrast with previous analyses
  (see {\it e.g.} Ref.~\cite{Leader:2014uua}) in which a tension
  between DIS and SIDIS data was claimed, with the former leading to a
  markedly negative total polarised strange PDF and the latter to a
  sign-changing one. We ascribe this finding to the flexibility of our
  parametrisation in conjunction with the usage of the {\sc MAPFF1.0}
  kaon FF set~\cite{AbdulKhalek:2022laj} which was determined with the
  same methodology used to determine the current PDFs.

\item The impact of theoretical constraints on the determination of
  polarised PDFs is variegated. On the one hand, the
  inclusion of data for $a_3$ and $a_8$ has negligible impact on both
  NLO and NNLO determinations: PDF central values and uncertainties
  remain almost unaffected. We therefore conclude that the current DIS
  and SIDIS data does not point towards any significant violation of
  the SU(2) and SU(3) flavour symmetries. The impact of the positivity
  constraint is instead very significant. While the quality of the
  NNLO fit improves when removing it, PDFs display peculiar shapes
  with large uncertainties in the large-$x$ region.
  
\end{itemize}

Our analysis can be improved on two main fronts. First, by including
hadronic data, specifically for gauge boson production, inclusive
hadron production, and single-inclusive and double-inclusive jet
production in polarised proton--proton collisions. The first process,
for which NNLO corrections are known~\cite{Boughezal:2021wjw}, can be
used to validate polarised sea quark PDFs obtained from SIDIS. The
other processes, being sensitive to the polarised gluon PDF starting
from LO, are expected to be instrumental in determining the
fraction of the proton spin carried by gluons. The second main front
concerns an improvement of the theoretical treatment by means of the
inclusion of theory uncertainties. In the spirit of recent analyses
carried out for unpolarised
PDFs~\cite{NNPDF:2019vjt,NNPDF:2019ubu,NNPDF:2024dpb,Ball:2020xqw},
these are related to both missing higher-order corrections in the
perturbative expansion and to nuclear corrections. The former may
help assess the relevance of the NNLO corrections and to estimate the
residual impact of contributions beyond NNLO. The latter may help
understand how good the isospin approximation is when relating
deuteron and proton structure functions.

In summary, our NNLO polarised PDF sets could be used in a
number of applications that require a matching theoretical
accuracy. For example, they can be employed to obtain accurate
predictions for DIS and SIDIS cross sections to be measured at the
future EIC, or they could serve as a baseline for the parametrisation
and the determination of polarised transverse-momentum-dependent
distributions at high accuracy.

\vspace{1cm}

\noindent The results presented in this paper have been obtained with the
public code~\cite{valerio_bertone_2024_10933177} available at
\begin{center}
  \href{https://github.com/MapCollaboration/Denali}{https://github.com/MapCollaboration/Denali},
\end{center}
with which we deliver our NLO and NNLO baseline polarised PDF sets in the
{\sc LHAPDF} format~\cite{Buckley:2014ana}. They are obtained from the
a global data set that includes DIS and SIDIS measurements as well as
the data for $a_3$ and $a_8$, and they obey the positivity constraint
in Eq.~\eqref{eq:pos_net}. The names of these PDF sets are:
\begin{itemize}
\item NLO: {\tt MAPPDFpol10NLO};
  \item NNLO: {\tt MAPPDFpol10NNLO}.
\end{itemize}
They will also be released on the {\sc LHAPDF} public
repository. The variant PDF sets discussed in
Sects.~\ref{subsec:data}-\ref{subsec:theory} are available from the
authors upon request.

\section*{Acknowledgments}

We thank Gunar Schnell for clarifications on the HERMES SIDIS data, Rabah
Abdul Khalek for contributions to the development of the numerical framework
used in this analysis, Leonardo Bonino and Giovanni Stagnitto for help
with the exact NNLO corrections to SIDIS matrix elements, and the members of
the MAP Collaboration for comments on the draft. V.~B. is supported by the
European Union’s Horizon 2020 research and innovation programme under grant
agreement STRONG 2020 - No 824093. E.R.~N. is supported by the Italian Ministry
of University and Research (MUR) through the “Rita Levi-Montalcini” Program.

\bibliography{MAPPolNNLO}

\end{document}